\documentclass{aa} 

\usepackage{graphicx}
\usepackage{txfonts}
\usepackage{natbib}
\usepackage{float}

\begin{document}

\title{The warm gas atmosphere of the HD 100546 disk seen by Herschel\thanks{{\it Herschel} is an ESA space observatory with science instruments provided by European-led Principal Investigator consortia and with important participation from NASA.}}

\subtitle{Evidence of a gas-rich, carbon-poor atmosphere?}

\author{Simon Bruderer\inst{\ref{inst_mpe},\ref{inst_eth}} \and Ewine F.~van~Dishoeck\inst{\ref{inst_leiden},\ref{inst_mpe}} \and Steven D.~Doty\inst{\ref{inst_denison}} \and Gregory J.~Herczeg\inst{\ref{inst_mpe}}}

\institute{
Max Planck Institut f\"{u}r Extraterrestrische Physik, Giessenbachstrasse 1, 85748 Garching, Germany\label{inst_mpe}
\and
Institute of Astronomy, ETH Zurich, 8093 Zurich, Switzerland\label{inst_eth}
\and
Leiden Observatory, Leiden University, PO Box 9513, 2300 RA Leiden, The Netherlands\label{inst_leiden}
\and
Department of Physics and Astronomy, Denison University, Granville, OH 43023, USA\label{inst_denison}
}

\date{Submitted on 7 October 2011 / Accepted to A\&A on 5 January 2012}

\titlerunning{The warm gas atmosphere of the HD100546 disk}
\authorrunning{Bruderer et al.}

\offprints{Simon Bruderer,\\ \email{simonbruderer@gmail.com}}


\abstract
{With the Herschel Space Observatory, lines of simple molecules (C$^+$, O, and high-$J$ lines of CO, $J_{\rm up}\gtrsim14$) have been observed in the atmosphere of protoplanetary disks. When combined with ground-based data on [\ion{C}{I}], all principle forms of carbon can be studied. These data allow us to test model predictions for the main carbon-bearing species and verify the presence of a warm surface layer. The absence of neutral carbon [\ion{C}{I}], which is predicted by models to be strong, can then be interpreted together with ionized carbon [\ion{C}{II}] and carbon monoxide.} 
{We study the gas temperature, excitation, and chemical abundance of the simple carbon-bearing species C, C$^+$, and CO, as well as O by the method of chemical-physical modeling. Using the models, we explore the sensitivity of the lines to the entering parameters and constrain the region from which the line radiation emerges.} 
{Numerical models of the radiative transfer in the lines and dust are used together with a chemical network simulation and a calculation of the gas energetics to obtain the gas temperature. We present our new model, which is based on our previous models but includes several improvements that we report in detail, together with the results of benchmark tests.} 
{A model of the disk around the Herbig Be star HD 100546 is able to reproduce the CO ladder together with the atomic fine-structure lines of [\ion{O}{I}] and either [\ion{C}{I}] or [\ion{C}{II}]. We find that the high-$J$ lines of CO can only be reproduced by a warm atmosphere with $T_{\rm gas} \gg T_{\rm dust}$. The low-$J$ lines of CO, observable from the ground, are dominated by the outer disk with a radius of several 100 AU, while the high-$J$ CO observable with Herschel-PACS are dominated from regions within some tens of AU. The spectral profiles of high-$J$ lines of CO are predicted to be broader than those of the low-$J$ lines. We study the effect of several parameters including the size of the disk, the gas mass of the disk, the PAH abundance and distribution, and the amount of carbon in the gas phase.} 
{The main conclusions of our work are \textit{(i)} only a warm atmosphere with $T_{\rm gas} \gg T_{\rm dust}$ can reproduce the CO ladder. \textit{(ii)} The CO ladder together with [\ion{O}{I}] and the upper limit to [\ion{C}{I}] can be reproduced by models with a high gas/dust ratio and a low abundance of volatile carbon. These models however produce too small amounts of [\ion{C}{II}]. Models with a low gas/dust ratio and more volatile carbon also reproduce CO and [\ion{O}{I}], are in closer agreement with observations of [\ion{C}{II}], but overproduce [\ion{C}{I}]. Owing to the uncertain origin of the [\ion{C}{II}] emission, we prefer the high gas/dust ratio models, indicating a low abundance of volatile carbon.} 

\maketitle

%
%

\section{Introduction} \label{sec:intro}

At an important phase of a young low-mass star's life, the natal envelope has dispersed, but the star is still surrounded by an optically thick protoplanetary disk, which is thought to be the location of planet formation (see \citealt{Dullemond10} and \citealt{Armitage10} for recent reviews). Our own solar system contains both gaseous giants and rocky planets. It is thus evident that both the dust and the gas content of the disk need to be studied in order to understand planet formation. The feedback of the forming star onto the disk can help decide the disk's fate. The star irradiates the disk, heating both the gas and dust and changing the chemical composition of the gas by high-energy ultraviolet (UV) radiation and X-rays. Through heating of the disk's upper atmosphere, a disk-wind is driven, which may eventually disperse the disk (\citealt{Alexander06,Ercolano08,Gorti09}). The most abundant species tracing the ionizing radiation and warm temperature of a few 100 K are (ionized) atomic carbon (C, C$^+$), atomic oxygen (O), and high-$J$ lines of CO (with $J_{\rm up}\gtrsim14$). Apart from neutral carbon, which is accessible to ground-based telescopes, these species can now be observed using the PACS instrument (\citealt{Poglitsch10}) onboard the Herschel Space Observatory (\citealt{Pilbratt10}).

The main forms of carbon (C, C$^+$, and CO) can thus be observed for the first time and through high-$J$ CO lines also in warm regions. This allows us to probe the carbon chemistry and the physical structure of planet- and comet-forming zones, which were previously inaccessible to us. The carbon budget is important for planet formation, since the terrestrial planets in the inner Solar System are known to have a deficit in carbon of three orders of magnitude relative to the solar abundance (\citealt{Allegre01}). Indications of a carbon deficit are also found towards other planetary systems (\citealt{Jura08}). In the diffuse ISM, about $40-60$ \% of the carbon is locked into amorphous carbon grains (\citealt{Savage96,Sofia04,Sofia11}). To be consistent with the Earth's composition, these grains need to have been destroyed before planet formation. In comets, however, carbon grains have been detected, suggesting that grain destruction mechanisms play no role (\citealt{Lee10}).

What region of the disk does the far-infrared (FIR) C, C$^+$, CO and O lines trace? How does the partitioning between volatile and refractory carbon vary from clouds to disks? The simultaneous measurement of the species tracing a significant fraction of the disk may provide answers to these questions. In addition, besides the hard to observe H$_2$, these simple species form the most abundant constituents of the gas. Their lines are thus important tracers of the physical structure of the disk and allow us to finally test the predictions made by many existing physical-chemical models of disks. Neutral carbon [\ion{C}{I}], for example, is predicted to be strong by models with a range of gas masses and dust settling (\citealt{Jonkheid07}) but is not detected in observations of HD 100546 (\citealt{Panic10}) and CQ Tau (\citealt{Chapillon10}). In addition to the submillimeter lines discussed in our work, neutral carbon also has forbidden lines in the near-infrared (at 8729, 9826, 9852 \AA), which are predicted to have detectable strengths (\citealt{Ercolano09a}). To date, only a few detections towards pre-main-sequence stars have been reported (\citealt{Looper10}).

The physics and chemistry of gas in protoplanetary disks, which is irradiated by the central star, is similar to the interface of general molecular clouds, where nearby massive stars heat and photoionize the gas. Modeling the infrared (IR) response of gas irradiated by UV and X-rays has been developed in the context of these photo-dominated or dissociation regions (PDRs) and X-ray dominated regions (XDRs) (e.g. \citealt{Tielens85,Sternberg89,Kaufman99,Meijerink05a}). These models simulate the infrared line emission by solving the coupled system of chemical evolution and thermal balance. The coupling is the  result of the cooling rates for the thermal balance depending on the abundance of the coolants, which is determined by the chemical network that itself also depends on the temperature. These physical-chemical models were applied in the context of protostars by \cite{Spaans95} and \cite{Bruderer09b,Bruderer10} to the walls along outflows, irradiated by UV radiation arising from either accretion hot-spots or the protostar's photosphere. For protoplanetary disks, physical-chemical models have been applied by e.g. \citet{Kamp01}, \citet{Glassgold04}, \citet{Kamp04}, \citet{Gorti04}, \citet{Nomura05}, \citet{Aikawa06}, \citet{Jonkheid04}, \citet{Jonkheid06}, \citet{Jonkheid07}, \citet{Gorti08}, \citet{Woitke09}, \citet{Woods09}, and \citet{Ercolano09b}.

The well-studied intermediate-mass pre-main-sequence star HD 100546 is the focus of our present study. The B9.5Vne type star of 2.2 $M_\odot$, $L_{\rm bol} \sim 27$ $L_\odot$, and $T_{\rm eff}=10500$ K (\citealt{vdAncker97}) is at its distance of $103\pm6$ pc (Hipparcos) one of the closest Herbig Ae/Be sources. The star and disk have been extensively observed at all wavelengths in lines and the continuum. X-ray observations are reported in \cite{Stelzer06}. Both FUSE and IUE observed HD100546 in the UV, detecting absorption in electronic transitions of H$_2$ probing the warm gas with a temperature of several 100 K (\citealt{MartinZaidi08}). \cite{Ardila07} analyzed scattered-light images in the optical using ACS/Hubble finding evidence of minimal grain sizes larger than those of typical ISM grains. The scattered light observations also reveal structures resembling spiral arms possibly caused by either the perturbations of a companion (\citealt{Quillen05}) or a warped disk structure (\citealt{Quillen06}). Features in the near-infrared (NIR) reveal a high crystalline silicate dust fraction in the inner disk surface layers (\citealt{Bouwman03}) and an abundant amount of PAHs whose emission is detected on extended scales (\citealt{Geers07a}). A wealth of continuum images in the near and mid-IR (e.g. \citealt{Pantin00,Grady01,Liu03,Grady05}) have helped to constrain the structure of the inner disk using SED fitting. They reveal a large inner hole of radius $\sim 10$ AU (\citealt{Malfait98,Bouwman03}), which possibly has a Jupiter-sized planet orbiting inside (\citealt{Acke06}). Interferometric observations (VLT-AMBER and MIDI, \citealt{Leinert04,Petrov07}) confirm the presence of the inner hole directly and set $r_{\rm hole} \sim 13$ AU (\citealt{Benisty10}). Some CO ro-vibrational emission at 4.7 $\mu$m was reported by \cite{vanderPlas09} and \cite{Brittain09}. The latter paper inferred that the CO emission originates from radii of 13-100 AU, which is consistent with the inner hole seen in the dust continuum. They found a hot rotational temperature of $\sim 1000$ K and indications that both UV fluorescence and collisional excitation are important for the excitation. Hot gas in the disk surface layers is also detected by observing the H$_2$ ${\rm v}=1-0$ S(1) line at 2.12 $\mu$m by \citet{Carmona11}.

The bulk of the disk mass is at cold temperatures, as traced by submillimeter lines and continuum. Submillimeter continuum was reported by \citet{Henning94,Henning98}, who derived a dust mass of a few times $10^{-4}$ $M_\odot$. The gas up to $\sim 100$ K is traced by CO rotational lines up to $J=7-6$, as reported by \cite{Panic10}. They derived a gas mass of $>10^{-3}$ $M_\odot$ and concluded from the high $^{12}$CO $J=6-5$/$^{12}$CO $J=3-2$ ratio that the disk atmosphere needs to be warmer than in T Tauri stars, probably owing to photoelectric heating of the gas by the strong UV field of the forming B9 star. Strong higher-$J$ lines up to CO $J=30-29$ were detected by \citealt{Sturm10} using the PACS instrument onboard Herschel. The detected lines have upper level energies of up to 2500 K.

In this work, we study the chemistry and excitation of simple species (C, C$^+$, CO, and O) in the disk around a pre-main-sequence Herbig Ae/Be star. Our approach is twofold. We first want to interpret the PACS observations of \citeauthor{Sturm10} (2010) and the ground-based observations of \citet{Panic10} taken towards HD 100546 and search for evidence of a warm disk atmosphere from the CO ladder, which probes a wide range of temperatures. We however also wish to study the dependence of the lines on certain parameters such as the carbon abundance in the gas phase, the gas/dust ratio, and others. In Section \ref{sec:methobs}, we briefly summarize the modeling approach and describe the observations. In Section \ref{sec:repmod}, we first present the results for one particular model that agrees reasonably well with the observations in Section \ref{sec:repmod} and then discuss the parameter dependences in Section \ref{sec:paramstud} and \ref{sec:discuss}. In Section \ref{sec:conclus}, we summarize our results.

%
%
\section{Model and observations} \label{sec:methobs}

For this work, we use a new model based on \citet{Bruderer09a,Bruderer09b,Bruderer10}, with several changes and improvements implemented. The section provides a short summary of the model. Details and results of benchmark tests are reported in Appendix \ref{sec:app_detail}.

Our physical-chemical model starts with \emph{(i)} a dust radiative transfer calculation for a given dust and gas distribution to obtain the local UV radiation field and the dust temperature. Using an initial guess of the gas temperature, \emph{(ii)} the chemical abundances are calculated, which serve as input for a molecular/atomic excitation calculation to obtain the cooling rates. Next, \emph{(iii)} the thermal balance is calculated to obtain an improved guess for the gas temperature. Steps \emph{(ii)} and \emph{(iii)} are repeated until convergence is achieved. Raytracing and convolution to the telescope beam then yields line fluxes that can be observed with observations. These modeling steps are summarized in Figure \ref{fig:modflow}.

Differences from the study of \citet{Bruderer09a,Bruderer09b,Bruderer10} are that we solve the chemical network for the steady-state solution instead of interpolating from a pre-calculated grid of abundances (\citealt{Bruderer09a}). This new approach allows us to include detailed photodissociation and ionization rates based on the local wavelength-dependent UV field and the actual column densities for the self-shielding factors. The thermal balance is refined by obtaining atomic and molecular cooling rates from the excitation calculation used in \citet{Bruderer10}. This (approximate) method, which is similar to that of \cite{Poelman05}, allows us to calculate the molecular excitation quickly and with reasonable accuracy. We can now include the detailed velocity structure into the excitation calculation and account for heating of the gas by the IR pumping of molecules, which is then followed by collisional deexcitation. The method couples cells at different radii of the disk, unlike the ``vertical direction only'' escape probability approach employed in most previous disk modeling. The disadvantage of our method is that we need to iterate over the entire chemical structure to derive the local gas temperature, which increases the calculation time. However, the method allows us to calculate models with an arbitrary geometry including e.g. disks embedded in envelopes. Since our code contains all geometrical properties in one module, it can be extended to three-dimensional structures in the future. The chemical network in the new code has also been extended by incorporating hydrogenation reactions on grain surfaces, photodesorption of species frozen-out onto grains, and vibrationally excited molecular hydrogen (H$_2^*$), which can be used to overcome activation barriers.

In this study, we concentrate on the rotational ladder of CO and do not consider vibration-rotation lines detected in the M-band at 4.7 $\mu$m, since fluorescent excitation by UV photons plays a significant role in their population. We note that the excitation of the rotational ladder is determined by collisions and the relative population of CO in the $\nu=1$ state is small at 4 \% for a thermalized population at 1000 K (\citealt{Brittain09}). Thus, the results presented in this work would not change by including vibrationally excited levels of CO. For the same reasons, we also do not model the H$_2$ 2.12 $\mu$m line. In addition, this line is very sensitive to the depth at which the dust becomes optically thick and a quantitative comparison to the far-infrared lines is thus difficult.

\subsection{Parameters for the HD 100546 model} \label{sec:modhd100}

To model HD 100546, we chose to use the dust density distribution obtained by \citet{Mulders11}. They fit the SED using a dust radiative transfer calculation with a vertical structure that obeys hydrostatic equilibrium assuming that $T_{\rm gas}=T_{\rm dust}$. 
Dropping this assumption results in a density structure with a puffed-up inner rim with about 10 AU (\citealt{Kamp09}), but this region does not contribute significantly to the emission we probe here. For consistency, we also use the dust temperature of \citet{Mulders11}. The surface density power-law is taken to be $\Sigma \propto r^{-1}$, which fits the VISIR images. Another dust density structure for the HD 100546 disk was proposed by \cite{Benisty10}, based on a analytical disk structure and fitted to the IR interferometric data. Their main conclusions about the inner disk were incorporated into the \citet{Mulders11} model that we use. It is well-known that SED fitting yields degenerate solutions and the obtained dust structure does not necessarily reflect the gas structure, e.g. owing to dust settling effects. In addition, the structure of the disk is likely non-axisymmetric (\citealt{Panic10,Quanz11}) and with substructures (\citealt{Ardila07}). However, this approximate approach is a reasonable one for studying the main characteristics of the gas emission, given the large uncertainties in the gas line modeling (Section \ref{sec:uncerttgas}).

We adopted a Hipparcos distance of 103 pc to HD 100546 (\citealt{vdAncker98}). The inclination and position angle were taken to be $42^\circ$ and $145^\circ$, respectively (\citealt{Pantin00}). We used a Keplerian velocity field around a 2.5 $M_\odot$ star, which was found by \citet{Panic10} to closely reproduce the observed spectra. The outer radius is uncertain but estimated to be between 200 AU and 400 AU (\citealt{Pantin00,Panic10}).

Inside the large inner hole of the HD 100546 disk, a smaller inner disk extending from 0.25 AU to 4 AU was found (\citealt{Benisty10}). We run models including the inner disk, but found that more than 99 \% of the flux of the lines discussed in this work emerges from the outer disk. We thus considered only the outer disk starting at 13 AU for this work. The dust opacity of the inner disk was however included to calculate the local UV field.

\begin{figure}[htb]
\sidecaption
\includegraphics[width=1.0\hsize]{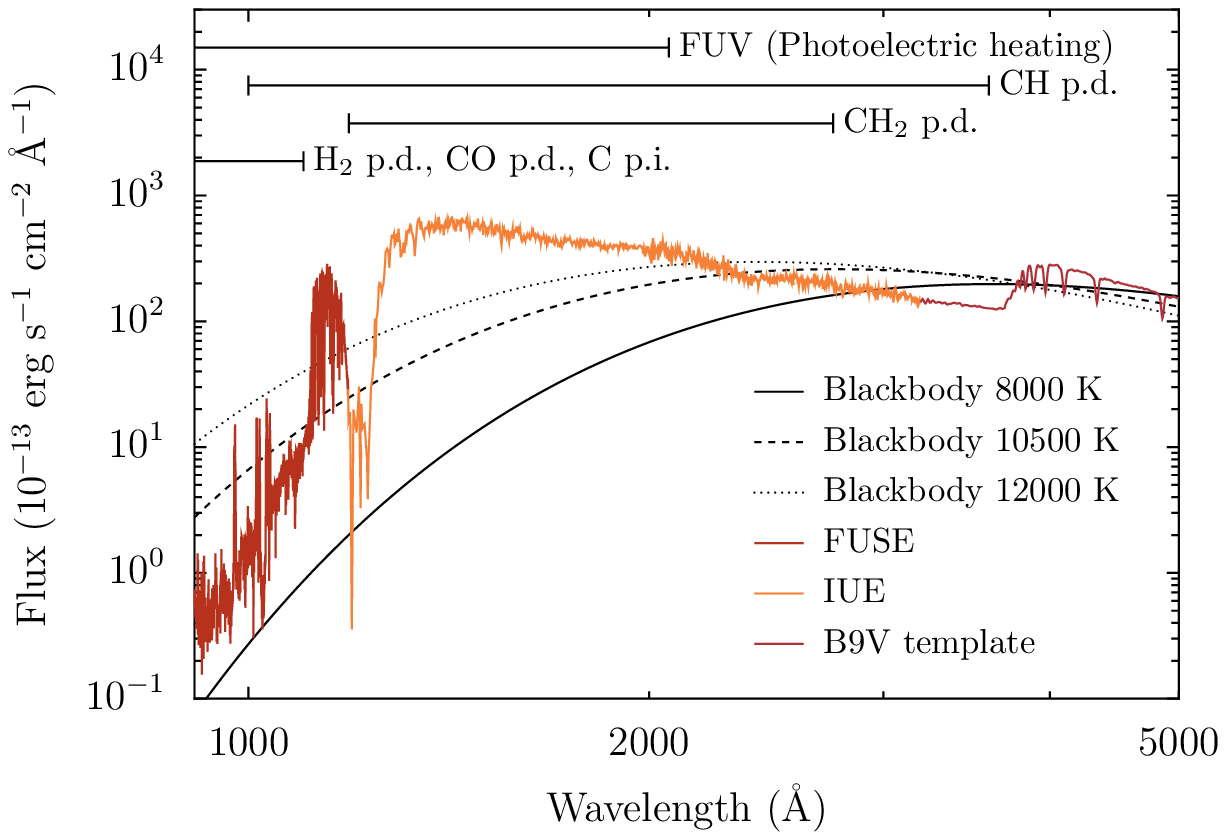}
\caption{UV spectra of HD 100546 at the source distance of 103 pc. We indicate the far ultraviolet (FUV)  wavelength range important for gas heating and the wavelength range of photodissociation (p.d.) and photoionization (p.i.) cross-sections for different molecules.}
\label{fig:ste_uv_spec}
\end{figure}

As input stellar radiation field (Figure \ref{fig:ste_uv_spec}), we analyzed dereddened FUSE observations for wavelengths 911-1200 \AA, which had been previously analyzed by \citet{Deleuil04} and IUE observations (\citealt{Valenti03}) for longer wavelengths (1200-3200 \AA). The spectra were dereddened using the extinction of $A_V=0.36$ (\citealt{Ardila07}) and the extinction law of \cite{Draine03b} for $R_{\rm V}=3.1$. The FUSE/IUE spectra were extended to longer wavelengths using the B9V template of \citet{Pickles98}. The X-ray luminosity had been found to be $\log(L_{\rm X}) = 28.9$ erg s$^{-1}$ (\citealt{Stelzer06}). Since the plasma temperature had not been well-constrained by observations, we assumed a rather high temperature of $k T_{\rm X}=6$ keV. This hard spectra allows X-rays to penetrate deep into the atmosphere. Nevertheless, the lines studied in this work change by less than $\sim 5$ \% by switching on X-rays. Thus, the assumption on the shape (plasma temperature) of the X-ray spectra does not affect our results (Section \ref{sec:physstruc}). In addition to the stellar radiation field, we also account for the interstellar radiation field (\citealt{Draine78}). Another source of ionization are cosmic rays, for which a rate of $\zeta_{\rm c.r.}=5 \times 10^{-17}$ s$^{-1}$ is adopted.

\begin{figure}[htb]
\includegraphics[width=1.0\hsize]{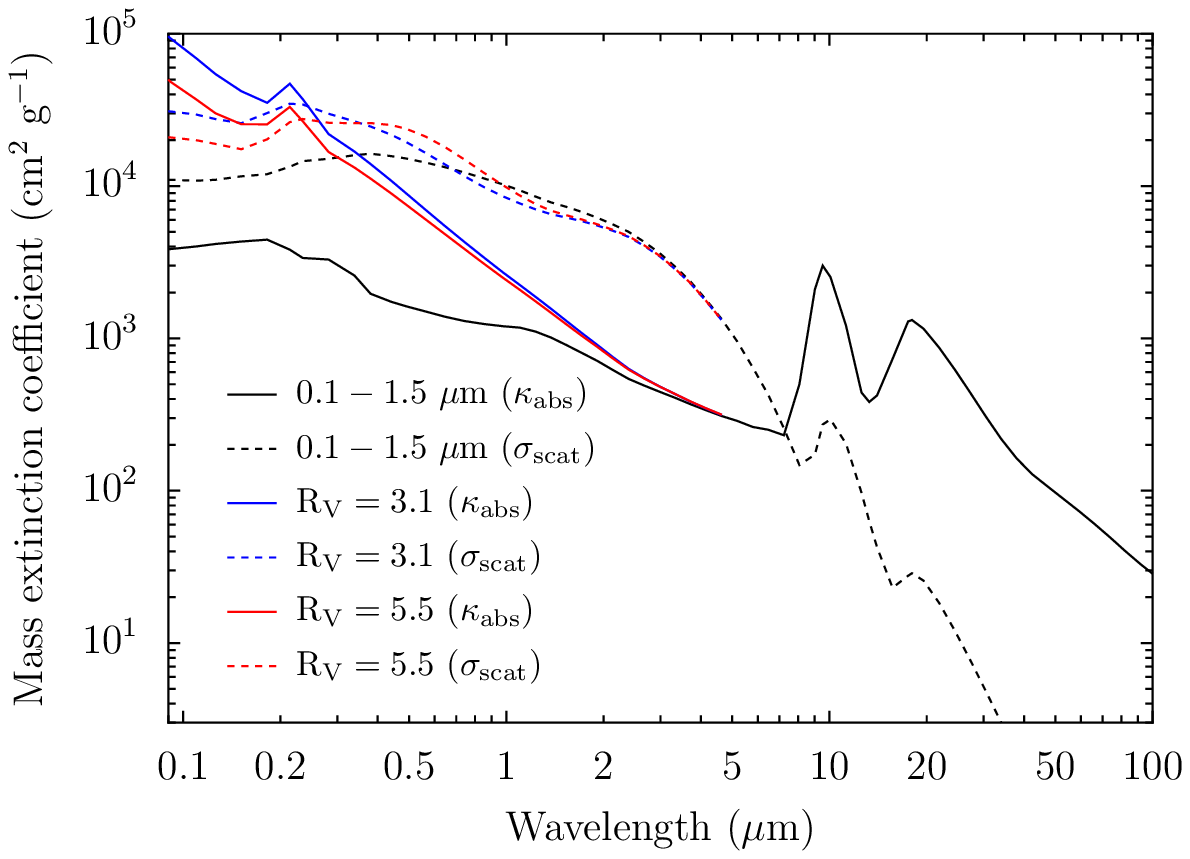}
\caption{Adopted dust opacities. The solid lines give absorption mass extinction coefficients, while dashed lines indicated scattering mass extinction coefficients.}
\label{fig:dustopac}
\end{figure}

The dust opacities control the penetration of the photons and thus both the photodissociation and ionization of molecules as well as the dust temperature. The dust grains used in the SED fitting by \citet{Mulders11} are silicates with a 5 \% (in mass) composition of carbonaceous material. For gas/dust $=100$ and a present-day solar photosphere carbon abundance of $2.7 \times 10^{-4}$ relative to hydrogen (\citealt{Asplund09}), this amounts to about 15 \% carbon bound in this population of grains. The grains have a size of 0.1 - 1.5 $\mu$m with a distribution proportional to $a^{-3.5}$, employing the Mathis, Rumpl and Nordsieck size distribution (\citealt{Mathis77}). The optical constants are for irregularly shaped distributions of hollow spheres (DHS, \citealt{Min05}). The DHS dust model has been adopted to be consistent with the SED fitting of \citet{Mulders11}. This dust population has a mass of $10^{-4}$ $M_\odot$, which agrees with observations of small grains by \citet{Dominik03}. However, much of the mass can be in larger grains, and the gas/dust ratios given relative to this population of grains are thus upper limits. This population of grains does not contain the UV/visual attenuation by very small grains (VSGs) and polycyclic aromatic hydrocarbons (PAHs). We thus also use $R_{\rm V}=3.1$ and 5.5 opacities after \citet{Draine03b} at wavelengths $\lesssim 1$ $\mu$m (Figure \ref{fig:dustopac}) to include VSG/PAH absorption. Using these opacities at short wavelengths only affects (i.e. increases) the dust temperature in the uppermost layer, above the region from which the line emission discussed in this work emerges, as we checked. For simplicity, we thus use the same dust temperature for all models taken from \citet{Mulders11}.

The PAH abundance affects both the heating by means of the photoelectric effect, as well as the ionization balance, by means of recombination reactions. We adopt a 5 \% PAH-to-dust ratio within 100 AU and 1\% outside 100 AU, following a simultaneous fitting of Spitzer and VISIR N-band observations using the $N_C=100$ opacities of \citet{Draine07} (Gijs Mulders, pers. comm.). The PAH-to-dust mass ratio of 5 \% corresponds approximately to the average Galactic PAH-to-dust mass ratio of 4.6 \%, which in turn represents about 50 ppm of C per H nucleon bound in PAHs (\citealt{Draine07}) and 15 \% of the present-day solar photosphere carbon abundance (for gas/dust $=100$). The PAHs are only excited by UV in the uppermost layer of the disk (\citealt{Visser07}) and the observed features thus trace this region. In addition to the large uncertainties in the molecular properties of the PAHs, the PAH abundance is only poorly determined in the region from which the lines discussed in this study emerge. We thus choose to scale the PAH abundance with the gas/dust ratio, such that gas with gas/dust $=20$ contains five times smaller amount of PAHs than for gas/dust $=100$ (\citealt{Jonkheid07}). We examine the influence of the PAH abundance separately in Section \ref{sec:parampah}. We assume PAHs to be photodestroyed in regions with integrated far ultraviolet (FUV) fluxes $>10^6$ ISRF, following the results of a more complete PAH chemistry by \citet{Visser07}. This affects however only the upper layer of the disk and we find that the lines discussed in this work are unaffected by this assumption. We note that the emission of all lines studied here emerges from $z/r$ lower than the region of PAH destruction (Section \ref{sec:physstruc}, \ref{sec:lineorigin}).

\subsection{Observations} \label{sec:observ}

Spectrally resolved observations of the low-$J$ $^{12}$CO $J=7-6$, $J=6-5$, and $J=3-2$ lines were obtained with the 12m APEX telescope by \citet{Panic10} and displayed a double-peaked line shape, that is consistent with a rotating disk. We note that the uncertainty in the $J=7-6$ line is considerable and the $J=6-5$ line is detected at a much higher signal-to-noise ratio. They have also searched for the 370 $\mu$m (809 GHz) [\ion{C}{I}] ${}^3P_2 - {}^3P_1$ line, but were unable to detect it. We note that they have been using position switching with an off position 16$\arcmin$ from the source. Thus, it is unlikely that the emission at the source had been canceled out by extended emission. The 609 $\mu$m (492 GHz) ${}^3P_1 - {}^3P_0$ line of neutral carbon was not observed. 

\begin{table}[htb]
\centering
\caption{Line fluxes measured towards HD 100546.}\label{tab:flux}
\begin{tabular}{llrr} 
\hline\hline
Species & Line & Wavelength\tablefootmark{a} & Flux/Error\tablefootmark{b} \\
& & [$\mu$m] & $10^{-17}$ [W m$^{-2}$]  \\
\hline
\multicolumn{4}{l}{APEX (\citealt{Panic10})}  \\
CO &$J=3-2$ & 866.96 &        0.0149 $\pm$ 0.003 \\
CO &$J=6-5$ & 433.56 &        0.132 $\pm$ 0.007 \\
CO &$J=7-6$ & 371.65 &        0.11 $\pm$ 0.02 \\
\lbrack\ion{C}{I}\rbrack &${}^3P_{2}-{}^3P_{1}$ & 370.42 & $<$0.009 \\ 
\hline
\multicolumn{4}{l}{Herschel PACS (\citealt{Sturm10})}  \\
CO &$J=14-13$ & 186.00 &        7.4 $\pm$ 0.9 \\
CO &$J=15-14$ & 173.63 &       11.5 $\pm$ 0.8 \\
CO &$J=16-15$ & 162.81 &        8.3 $\pm$ 0.9 \\
CO &$J=17-16$ & 153.27 &       10.5 $\pm$ 0.8 \\
CO &$J=18-17$ & 144.78 &        9.9 $\pm$ 1.0 \\
CO &$J=19-18$ & 137.20 &        8.9 $\pm$ 0.7 \\
CO &$J=20-19$ & 130.37 &        7.1 $\pm$ 0.7 \\
CO &$J=21-20$ & 124.19 &        5.9 $\pm$ 0.6 \\
CO &$J=22-21$ & 118.58 &        5.9 $\pm$ 0.9 \\
CO &$J=23-22$ & 113.46 &     $<$4.8 $\pm$ 1.2 \\
CO &$J=24-23$ & 108.76 &       10.6 $\pm$ 1.3 \\
CO &$J=26-25$ & 100.46 &       12.7 $\pm$ 2.4 \\
CO &$J=27-26$ & 96.77  &       10.2 $\pm$ 1.4 \\
CO &$J=28-27$ & 93.35  &       10.0 $\pm$ 1.0 \\
CO &$J=29-28$ & 90.16  &       11.8 $\pm$ 1.6 \\
CO &$J=30-29$ & 87.19  &        8.5 $\pm$ 1.0 \\
CO &$J=31-30$ & 84.41  &     $<$4.9 $\pm$ 1.5 \\
\lbrack\ion{C}{II}\rbrack &${}^2P_{3/2}-{}^2P_{1/2}$ & 157.74 & 13.5 $\pm$ 1.5\tablefootmark{c} \\ 
\lbrack\ion{O}{I}\rbrack &${}^3P_1 - {}^3P_2$ & 63.18 & 554 $\pm$ 5 \\
\lbrack\ion{O}{I}\rbrack &${}^3P_0 - {}^3P_1$ & 145.53 & 35.7 $\pm$ 1.3 \\
\hline
\end{tabular}
\tablefoot{\tablefoottext{a}{Rest frequency} \tablefoottext{b}{Formal errors, not including systematic errors} \tablefoottext{c}{Extended emission subtracted, see text.}}
\end{table}

The [\ion{C}{II}] 158 $\mu$m ${}^2P_{3/2}-{}^2P_{1/2}$, the [\ion{O}{I}] 63 $\mu$m ${}^3P_1 - {}^3P_2$,  and 145 $\mu$m ${}^3P_0 - {}^3P_1$ lines and mid-$J$/high-$J$ lines of CO with $J_{\rm up} = 14$ to 30 were observed by \citet{Sturm10} with the PACS instrument onboard Herschel as part of the ``Dust, ice, and gas in time'' (DIGIT) key program (PI: N. Evans). The CO $J=31-30$ line was blended with the OH $3/2, 7/2^+ - 5/2^-$ line and CO $J=23-22$ was potentially blended with H$_2$O $4_{14}-3_{03}$. These two lines are thus taken as upper limits. We note that the apparent jump in CO line flux at $J>25$ may have been introduced by uncertainties in the PACS flux calibration when the first PACS results were published, including those of \citet{Sturm10}. The larger uncertainty in these lines does however not affect any of our conclusions. The PACS data are spectrally unresolved. Except for [\ion{C}{II}], all lines are consistent with their emergence from within the central $9\farcs4$ pixel. To correct for extended emission in [\ion{C}{II}], we subtracted the pixels neighboring the central pixel. We also accounted for emission leaking from the center pixel owing to the extended point spread function. The origin of the [\ion{C}{II}] line is discussed in Section \ref{sec:cIIemission}.

%
%
\section{Results: A representative model} \label{sec:repmod}

We now present the details of a \emph{representative} model of HD 100546. By representative, we mean that it reproduces sufficiently well the observations enabling us to study the main physical and chemical effects. It is however not meant to be a \emph{best fit} model. The dependence on the parameters assumed here is discussed in the next section. The outer radius of the representative model is assumed to be 400 AU. We define the depletion factor $\delta_{\rm C}$ as the carbon abundance in the gas phase relative to the present-day solar photosphere abundance of $2.7 \times 10^{-4}$ relative to hydrogen (\citealt{Asplund09}). In this section, we assume a carbon gas phase abundance of $1.3\times10^{-4}$ relative to hydrogen ($\delta_{\rm C} \sim 0.5$), which is characteristic of the diffuse ISM. Adopting gas/dust $=20$, we get a gas mass of $\sim 2 \times 10^{-3}$ $M_\odot$. We use the $R_V=5.5$ dust extinction law for this model.

\subsection{Physical structure} \label{sec:physstruc}

\begin{figure*}[htb]
\includegraphics[width=1.0\hsize]{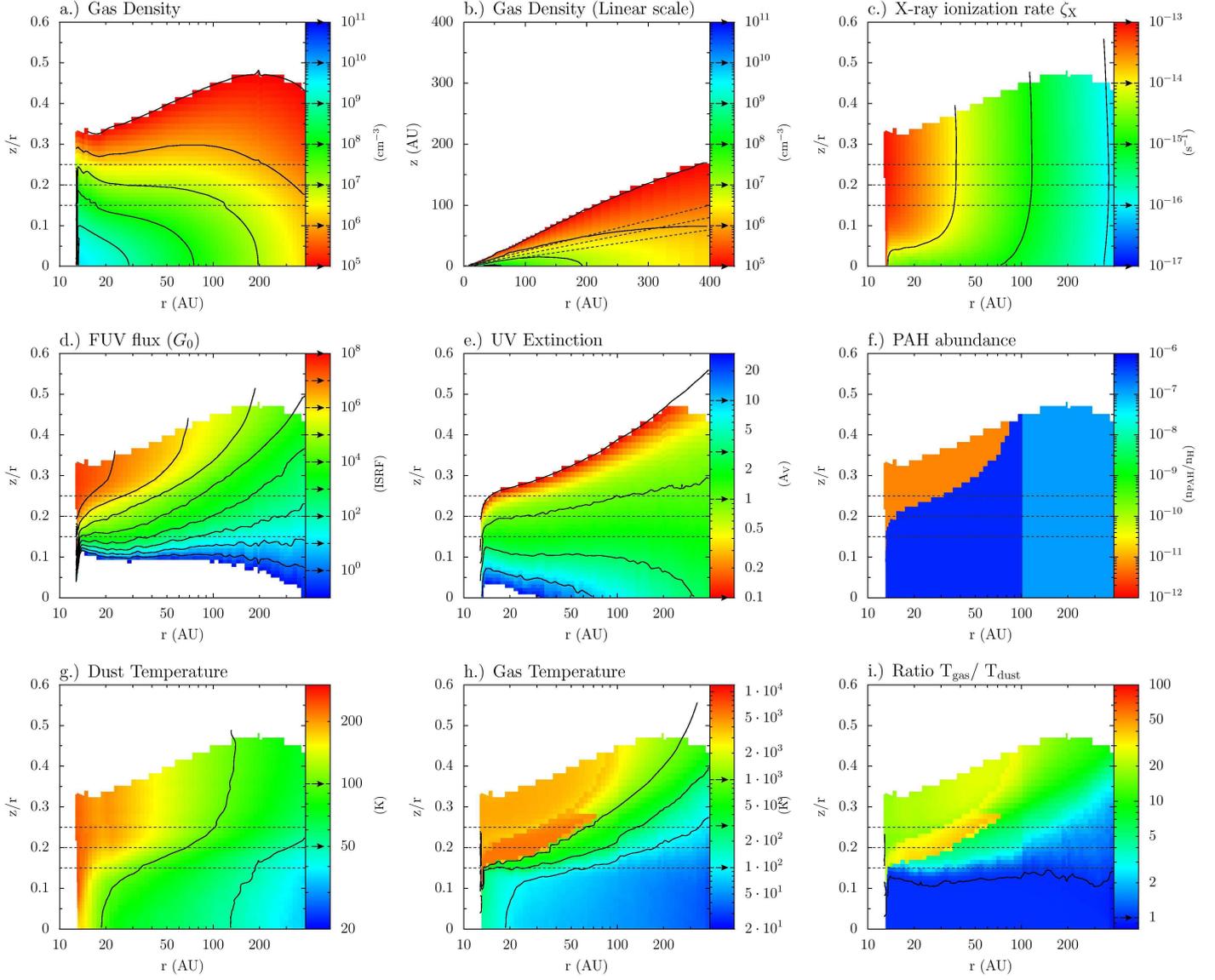}
\caption{Physical conditions in the representative model. The contour lines are shown for values indicated in the label by small arrows. The thin dashed line indicates $z/r=0.15,0.2$, and $0.25$. The panels show: \textbf{a.)} Gas density \textbf{b.)} Gas density in linear scale. \textbf{c.)} Total ionization rate. \textbf{d.)} Local FUV flux \textbf{e.)} FUV extinction \textbf{f.)} PAH abundance in units of the ISM abundance \textbf{g.)} Dust temperature \textbf{h.)} Gas temperature \textbf{i.)} Ratio of the gas to dust temperature.}
\label{fig:physstruct}
\end{figure*}

The physical conditions of the representative model are shown in Figure \ref{fig:physstruct}. Only one quadrant of the disk is given and only regions with gas density $>10^5$ cm$^{-3}$ are shown. We present the gas density in Figure \ref{fig:physstruct}a/b both in linear ($r$,$z$) as well as in ($\log(r)$,$z/r$)-space. The latter has the advantage that the inner part and upper disk can be seen clearly. Direct rays from the star to the disk correspond to vertical lines. The density reaches up to a few times $10^{10}$ cm$^{-3}$ in this model in the inner part of the disk midplane.

The X-ray ionization rate is given in Figure \ref{fig:physstruct}c. Despite the low X-ray luminosity of this source of $\sim 8 \times 10^{28}$ erg s$^{-1}$, the ionization rate due to X-rays in the inner disk is still about three orders of magnitude higher than the standard cosmic ray ionization rate of $5 \times 10^{-17}$ s$^{-1}$. For the heating and cooling budget of the disk, X-rays can however be neglected, since the luminosity in the FUV band (6-13.6 eV) is at $\sim 4 \times 10^{34}$ erg s$^{-1}$ much higher than the X-ray luminosity. Nevertheless, the considerable ionization rate may affect the chemistry and for example lead to a destruction of water (e.g. \citealt{Staeuber06}). The simple species considered in this work are however little affected by X-rays and we find that the CO, C, C$^+$, and O fluxes change by less than $\sim 5$ \% when X-rays are switched off.

The (integrated) FUV flux in units of the interstellar radiation field (\citealt{Draine78}; 1 ISRF $\sim 2.7 \times 10^{-3}$ erg s$^{-1}$ cm$^{-2}$) is shown in Figure \ref{fig:physstruct}d. Very high FUV fluxes of up to $10^8$ ISRF are reached in the surface of the inner edge of the disk. In the upper atmosphere of the outer disk at $r=400$ AU and $z/r=0.3$ the FUV flux still reaches $10^5$ ISRF at a density of a few times $10^5$ cm$^{-3}$. We note that these are roughly the conditions of the Orion bar PDR (\citealt{Hogerheijde95b,Jansen95b}). Similar conditions are also predicted for the outflow walls, the ``dense PDR'' along the outflow of a high-mass star forming region (\citealt{Bruderer09b}).

Figure \ref{fig:physstruct}e gives the FUV extinction in units of $A_{\rm V}$, obtained by defining $\tau_{\rm FUV}=-\log(I_{\rm att}/I_{\rm unatt})$, with the attenuated FUV flux $I_{\rm att}$ and the FUV flux corrected only for geometrical dilution $I_{\rm unatt}$ (e.g. \citealt{Bruderer09b}, Eq. 2). Conversion to $A_{\rm V}$ is done by scaling with the dust opacities at 2070 \AA~ and 555 \AA~ and accounting for $A_{\rm V}Ê\sim 1.086 \tau_{\rm V}$. The UV extinction is not used during the calculation, because the photoionization and dissociation rates are calculated from the wavelength-dependent UV spectra. The FUV extinction however shows that owing to scattering, FUV photons can penetrate much further into the disk, particularly in the outer parts (e.g. \citealt{vanZadelhoff03}).

Figure \ref{fig:physstruct}f shows the adopted PAH abundance. The drop in the inner, upper disk is due to the photodissociation of PAHs. The drop at 100 AU is observationally constrained (Section \ref{sec:modhd100}).

The dust and gas temperature are presented in Figure \ref{fig:physstruct}g/h. Figure \ref{fig:physstruct}i shows the ratio of the gas to dust temperature. The dust temperature does not exceed $\sim 250$ K, but stays above $\sim 30$ K across the entire model. Thus, CO does not freeze out onto the dust grains. The gas temperature exceeds the dust temperature by factors of up to 50 in the inner upper disk and reaches temperatures of up to $\sim 6000$ K. In the outer upper disk, the gas temperature is still $T_{\rm gas} \sim 1000$ K and thus about a factor of ten higher than $T_{\rm dust}$. Deeper within the disk, below $z/r = 0.15$, collisions between gas and dust grains result in a coupling of the gas and dust temperature ($T_{\rm gas}=T_{\rm dust}$). 

\subsection{Chemical structure} \label{sec:chemstruct}

\begin{figure*}[htb]
\includegraphics[width=1.0\hsize]{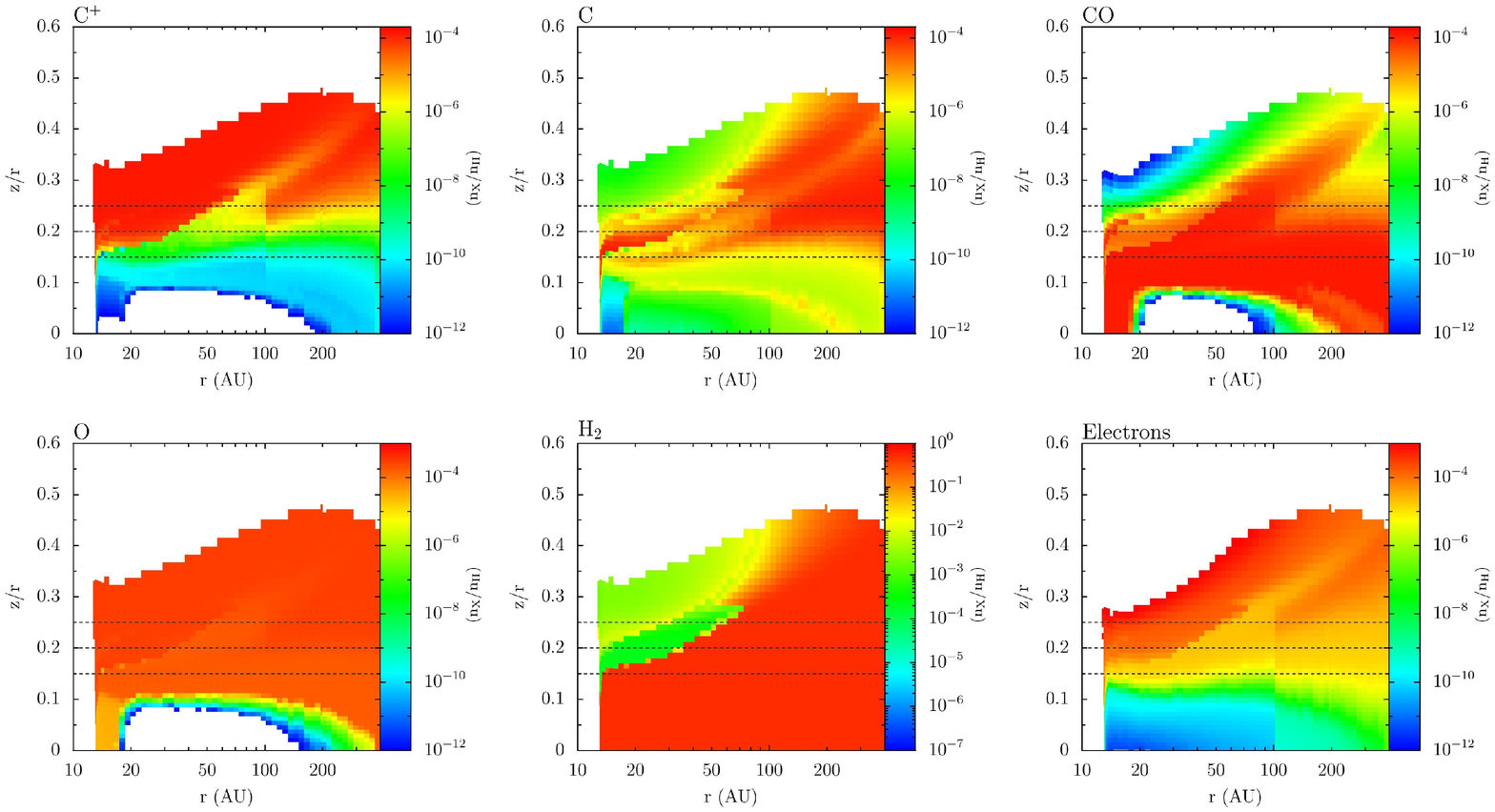}
\caption{Fractional abundances of C$^+$, C, CO, O, H$_2$, and electrons relative to the total hydrogen density ($n_{\rm H} = n({\rm H}) + 2n({\rm H}_2)$). The thin dashed line indicate $z/r=0.15,0.2$, and $0.25$. Abrupt changes at $R=100$ AU are due to a varying PAH abundance, see text.}
\label{fig:chemstruct}
\end{figure*}

Fractional abundances of C$^+$, C, CO, O, H$_2$, and electrons relative to the total hydrogen density ($n_{\rm H} = n({\rm H}) + 2n({\rm H}_2)$) are shown in Figure \ref{fig:chemstruct}. As found by previous chemical models of protoplanetary disks the chemical structure resembles a classical PDR in the vertical direction (e.g. \citealt{Jonkheid07}). Below a top layer of atomic hydrogen, H$_2$ forms, as self-shielding becomes sufficient. In the outer disk, the inward column is also large enough to prevent H$_2$ from photodissociation and hydrogen is in molecular form. 

Carbon in the upper disk ($z/r \sim 0.2$) is mainly in the form of C$^+$, followed by neutral carbon and CO deeper within the disk. An exception is the ``warm finger'' of CO in the outer disk ($r >$ 100 AU, $z/r > 0.25$). In this region, CO forms through the reaction of C$^+$ with H$_2$ to CH$^+$, followed by a reaction with O to form CO$^+$ which then reacts with H$_2$ to form HCO$^+$ and subsequently CO (see \citealt{Jonkheid07}). The initiating reaction is strongly endothermic with an energy barrier corresponding to 4640 K, thus requires a high temperature to proceed. The OH formation at high temperature, by means of the reaction of O and H$_2$, leads to additional CO$^+$ for this chain of reactions. The OH reacts with C$^+$ to form CO$^+$. At the inner edge of the disk at 13 AU below $z/r \sim 0.1$, CO is shielded from photodissociation owing to shadowing by the inner disk. The higher abundance of CO at $r=13$ AU and $z/r=0.2$ is due to the adopted H$_2$ formation rate: since the sticking coefficient decreases above $\sim 3000$ K (\citealt{Cuppen10}), the lower gas temperature at higher $z/r$ leads to an increase in H$_2$ and subsequently in CO.

Atomic oxygen has a considerably high abundance throughout the disk, except for the midplane. This is because the elemental abundance of oxygen is higher than that of carbon and not all oxygen can be locked into CO, which is almost unaffected by photodissociation, by means of self-shielding. In addition, oxygen cannot be ionized directly by FUV photons since its ionization energy is slightly higher than 13.6 eV. In the inner disk with $13$ AU $<r<20$ AU and $z/r < 0.1$, gas and dust are at a temperature $> 100$ K and the gas is not exposed to strong FUV radiation. In this part of the disk, oxygen is locked into water. Further out, water freezes out and locks most of the oxygen into water ice. In this region, the carbon is in the form of gaseous CH$_4$.

The electron fraction roughly follows the abundance of C$^+$. Charge exchange with PAHs is another important reaction type for the ionization balance. The PAHs are positively charged in the upper atmosphere, neutral at intermediate heights, and negatively charged at greater depths within the disk due to charging by electron attachment. Since the electron abundance is very low in the midplane, the PAHs can again become neutral. Charge exchange C$^+$ $+$ PAH$^0$ $\rightarrow$ C $+$ PAH$^+$ and recombination C$^+$ $+$ PAH$^-$ $\rightarrow$ C $+$ PAH$^0$ are the two reactions that affect the simple carbon chemistry most, leading a higher abundance of neutral carbon.

\subsection{Warm atmosphere versus dust temperature} \label{sec:warmatmos}

\begin{figure*}[htb]
\sidecaption
\includegraphics[width=0.7\hsize]{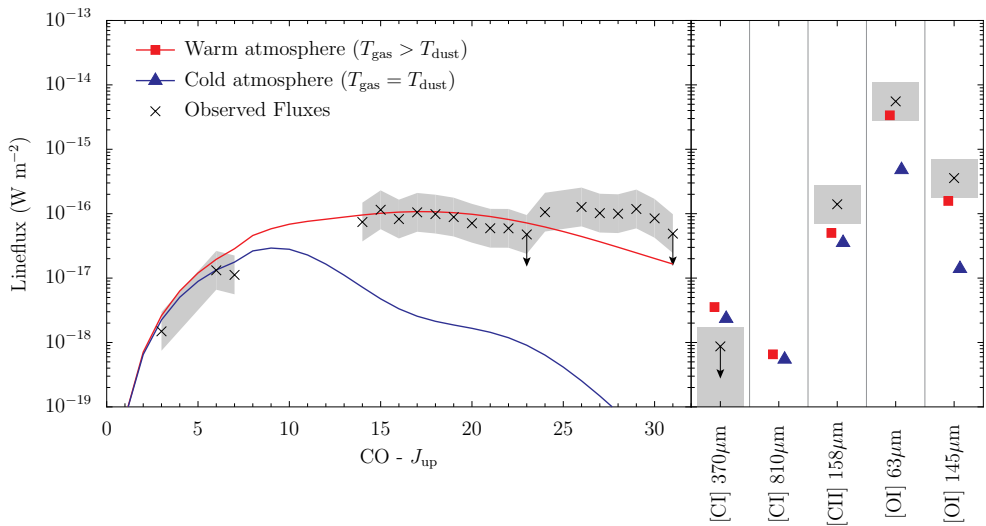}
\caption{Integrated line fluxes obtained from the representative model. The left part of the figure shows the CO ladder and the right part atomic fine-structure lines. Observations are indicated by black crosses. The red line and squares show model fluxes for a model with calculated gas temperature, the blue line and triangles with $T_{\rm gas}$ set to $T_{\rm dust}$. The gray shaded region indicates a factor of two to the observed fluxes.}
\label{fig:fluxreference}
\end{figure*}

Line fluxes predicted by the representative model are presented in Figure \ref{fig:fluxreference}. The figure gives the CO ladder on the left and the [\ion{C}{I}], [\ion{C}{II}], and [\ion{O}{I}] atomic fine-structure lines on the right. Atomic fine-structure lines of C and CO lines up to $J=7-6$ are convolved to the APEX beam and all other lines to the Herschel beam. In addition to the model with the calculated gas temperature, a model that assumes that the gas temperature is equal to the dust temperature is shown. A factor of two around the observed fluxes is shown by a gray shaded region. We consider this region as a good agreement to observations given all the uncertainties in data and modeling (see Section \ref{sec:uncerttgas}).

The model with the calculated gas temperature fits the data well, with the exception of the high-$J$ CO lines with $J > 25$ (see Sections \ref{sec:deph2} and \ref{sec:uncerttgas}). However, the cool atmosphere with $T_{\rm gas}$ set to $T_{\rm dust}$ fails to reproduce the high-$J$ CO ladder. The $J=16-15$ transition with $E_{\rm up} \sim 750$ K (Table \ref{tab:linorig}) is already underproduced by about an order of magnitude, and the higher-$J$ lines are underpredicted even more. For example,  $J=25-24$ ($E_{\rm up} \sim 1800$ K) is underpredicted by more than two orders of magnitude. The atomic fine-structure lines have upper level energies that are much lower than the high-$J$ lines of CO. Both the [\ion{C}{I}] and [\ion{C}{II}] lines have $E_{\rm up} < 100$ K and the line fluxes calculated with a cool and warm atmosphere are very similar. The [\ion{C}{II}] line is somewhat underproduced and [\ion{C}{I}] overproduced. However, [\ion{O}{I}] has an upper level energy of 327 K (145 $\mu$m) and 227 K (63 $\mu$m). Consequently, the [\ion{O}{I}] lines for the cool atmosphere differ by an order of magnitude from those obtained for the warm atmosphere. We conclude that the prominent emission in high-$J$ CO and [\ion{O}{I}] atomic fine-structure lines is strong evidence of a higher temperature of the gas in the upper atmosphere of the disk. This agrees to the recent detection of CH$^+$ (\citealt{Thi11}) and OH (\citealt{Sturm10}) towards this disk, which are both formed in endothermic reactions with H$_2$.

\subsection{Origin of the line emission} \label{sec:lineorigin}

\begin{figure*}[htb]
\includegraphics[width=1.0\hsize]{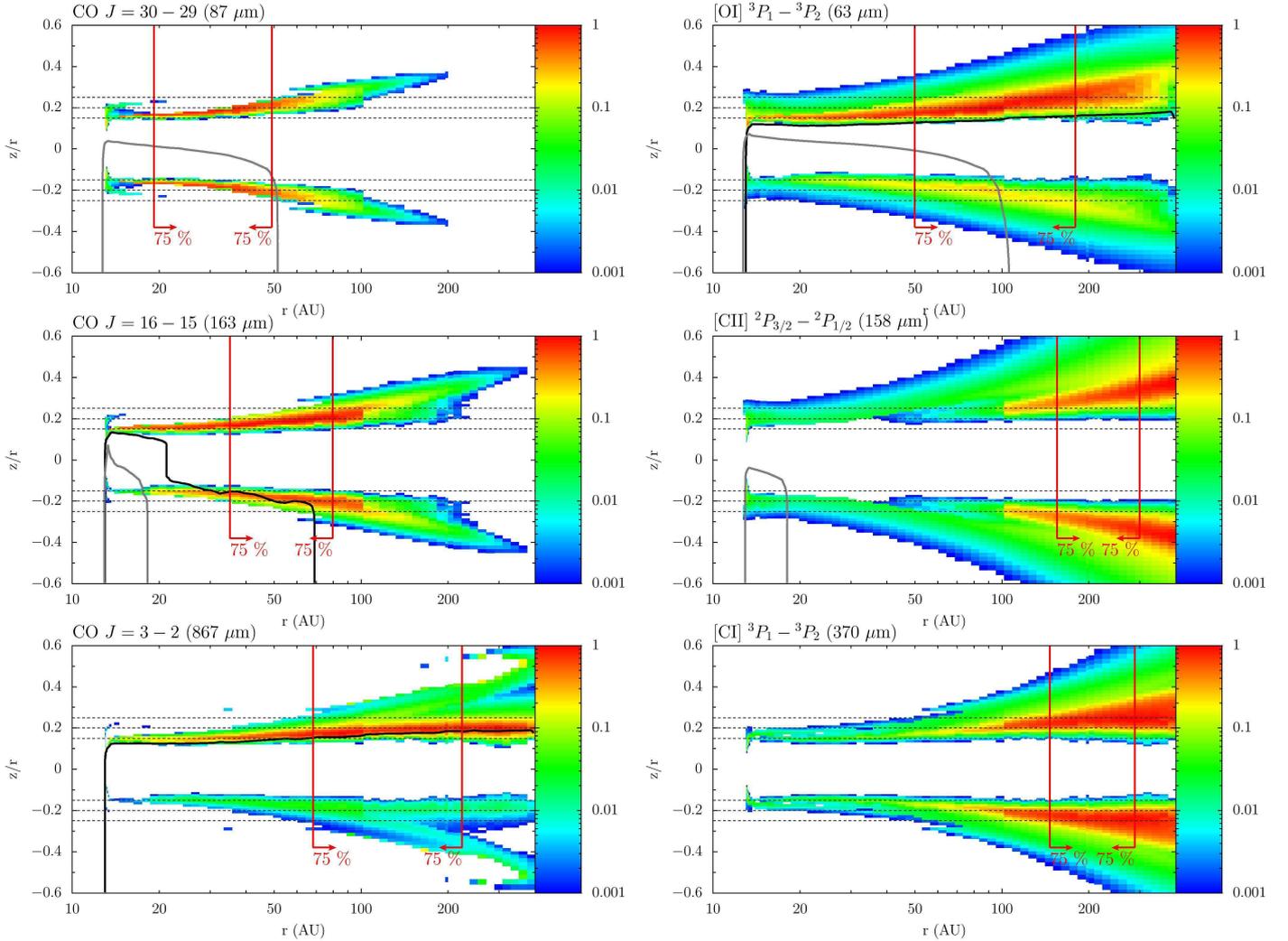}
\caption{Contribution function to the line formation. The disk is viewed from the top and the contribution function is normalized to the peak. Total (line+dust) opacities of $\tau_{\rm line center}=1$ at the line center are indicated by a black line, and dust opacities of $\tau_{\rm dust}=0.1$ by gray lines. The red arrows indicate the radii from which 75 \% of the emission emerges, either inside or outside. The thin dashed line indicates $\left|z/r\right|=0.15,0.2$, and $0.25$.}
\label{fig:contriplot}
\end{figure*}

Where does the observed line emission originate? The angular resolution of Herschel, corresponding to about 1000 AU for 63 $\mu$m at the distance of HD 100546, does not resolve the disk. The model however, may help us to locate the region of emission. Figure \ref{fig:contriplot} gives the relative contribution to the observed fluxes. We show the integrated contribution function (ICF), which incorporates all the effects involved in the line formation: abundance of the molecule, excitation, and opacity. In addition, we provide the two radii delineating the region from which 75 \% of the emission emerges, either inside or outside.

The ICF, discussed in \citet{Tafalla06} and \citet{Pavlyuchenkov08}, shows the relative contribution to the line emission along a ray. We obtain from the formal solution of the radiative transfer equation (Eq. \ref{eq:linert_solveradi})
\begin{equation}
{\rm ICF } = \int_{\rm line} (S_{\nu}-S_{\rm dust}) \ e^{-\tau_\nu} \ \left(1-e^{-\Delta\tau_\nu}\right) d\nu \ ,
\end{equation}
where the source function is $S_\nu$, the dust source function is $S_{\rm dust}$, the opacity to the observer is $\tau_\nu$ and the opacity within one cell is $\Delta\tau_\nu$. The integration is performed over frequency. For simplicity, we take the disk to be face on and allow the rays to propagate parallel to the $z$ axis (as e.g. in \citealt{Panic10}). This avoids complications with the adopted Keplerian velocity profile but conserves the main physical conclusions. The contribution to the observed line flux from an annulus located at a radial distance $r$ is given by $2\pi I(r) r dr$, with the flux $I(r)$ at radius $r$. Thus, the relative contribution to the observed flux is $\propto I(r) dr$. In Figure \ref{fig:contriplot}, we thus show the contribution function times the radius ($r \times {\rm ICF}$), scaled to its peak value.

The low-$J$ emission of CO $J=3-2$ originates at radii\footnote{Measured by the 75 \% contribution radii given in Figure \ref{fig:contriplot} and Table \ref{tab:linorig}.} between $r \sim 70 - 220$ AU at $z/r \sim 0.15-0.25$. The density above this region is too low to contribute significantly to the emission, since the regions below are shielded by the line opacity. Thus, the far side of the disk does not contribute much to the emission, except for some radiation in line wings. Owing to the increasing density and temperature towards the inner region at a given $z/r$, even spatially smaller regions at radii $< 100$ AU contribute to the observable emission. Mid-$J$ lines, e.g. CO $J=16-15$ with $E_{\rm up} \sim 750$ K, are not excited in the outer disk and only molecules within $r \sim 35 - 80$ AU contribute to the emission. These lines reach line opacities of around unity and the far side of the disk does not contribute to the emission in the innermost regions (radii $<30$ AU). The highest-$J$ emission, e.g. CO $J=30-29$ with $E_{\rm up} \sim 2800$ K, detected towards HD 100546 emerges from the inner $r \sim 20 - 50$ AU in a thin layer at the surface of the disk. There, CO is formed but the gas temperature is still sufficiently high to excite the lines. The opacity of these lines is optically thin and the dust extinction is not large. Thus, emission from both the near and far sides reaches the observer.

The atomic fine-structure lines of [\ion{C}{I}] and [\ion{C}{II}] emerge mainly from the outer disk at radii $\sim 150-300$ AU. Owing to the higher abundance of those species in the upper atmosphere, combined with the low critical density ($\sim 10^3$ cm$^{-3}$) and low upper level energy of the [\ion{C}{I}] and [\ion{C}{II}] lines, these lines emerge from regions higher in the disk than the low-$J$ lines of CO. The oxygen [\ion{O}{I}] lines have considerably higher critical densities of $\sim 10^5-10^6$ cm$^{-3}$ and also higher upper level energies. They are thus not excited in the outer disk and the line emission comes from regions within $\sim 50-210$ AU. Since this line is optically thick, most of the emission is from the near side of the disk.

\begin{table}[htb]
\centering
\caption{Line properties and origins}\label{tab:linorig}
\begin{tabular}{lrrrr|c} 
\hline\hline
Line & $\lambda$ & $E_{\rm up}$ & $A_{\rm ul}$& $n_{\rm crit}$\tablefootmark{a} & Origin \\
	& [$\mu$m] & [K] & [s$^{-1}$] & [cm$^{-3}$] & [AU] \\
\hline
CO\, $J=3-2$ & 867 & 33 & 3(-6) & 2(4) & 70-220 \\
CO\, $J=16-15$ & 163 & 752 & 4(-4) & 1(6) & 35-80 \\
CO\, $J=30-29$ & 87 & 2565 & 2(-3) & 4(6) & 20-50  \\
\lbrack\ion{C}{I}\rbrack\, ${}^3P_1 - {}^3P_2$ & 370 & 62 & 3(-7) & 1(3) & 150-300 \\
\lbrack\ion{C}{II}\rbrack\, ${}^2P_{3/2}-{}^2P_{1/2}$ & 158 & 91 & 2(-6) & 5(3) & 150-300 \\
\lbrack\ion{O}{I}\rbrack\, ${}^3P_1 - {}^3P_2$ & 63 & 228 & 9(-5) & 5(5) & 50-180 \\
\lbrack\ion{O}{I}\rbrack\, ${}^3P_0 - {}^3P_1$ & 145 & 327 & 2(-5) & 5(4) & 80-210 \\
\hline
\end{tabular}
\tablefoot{a(b) means $a\times10^b$. \tablefoottext{a}{Critical density ($A_{ul} / \sum_l C_{ul}$) for collision with H$_2$ at the temperature of $E_{\rm up}$.} Atomic/molecular data from \cite{Schoier05}.}
\end{table}

\subsection{CO line profile} \label{sec:colineshape}

\begin{figure}[htb]
\includegraphics[width=1.0\hsize]{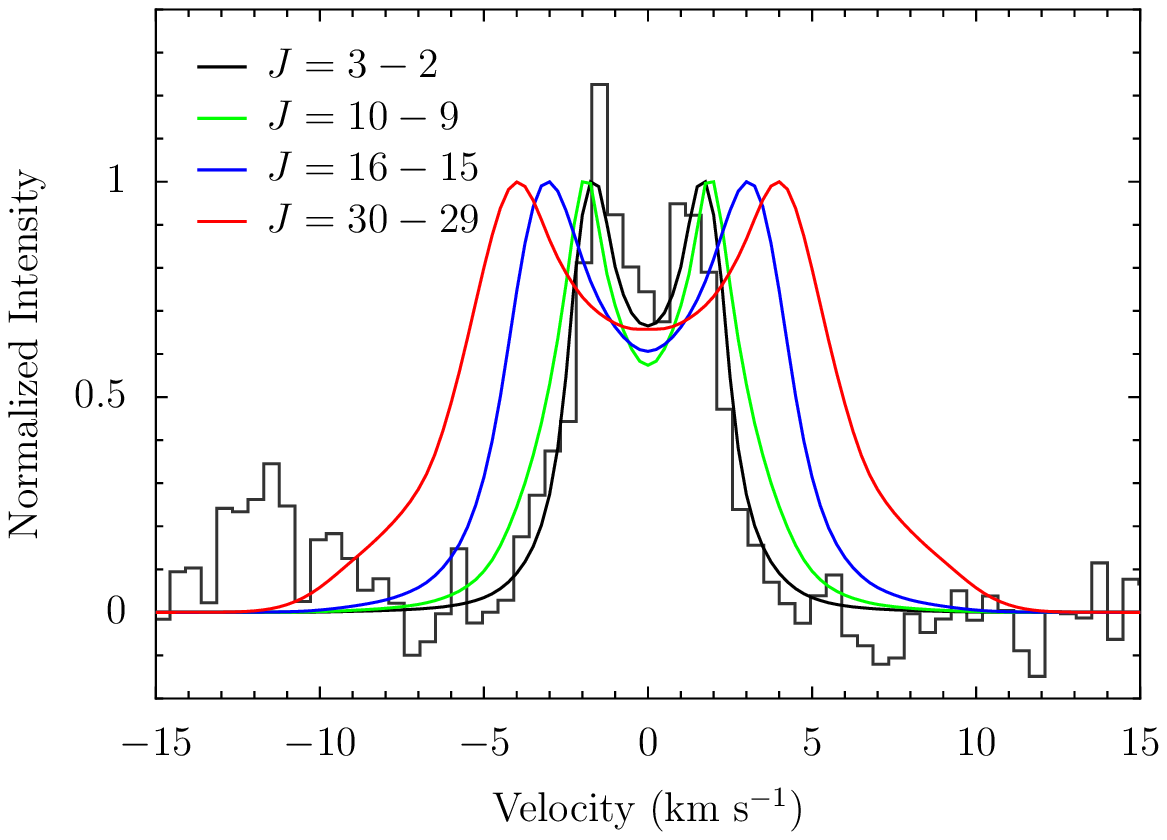}
\caption{Predicted line shapes for the CO $J=10-9$, $J=16-15$, and $J=30-29$ in the beam of Herschel. The CO $J=3-2$ line predicted for APEX is shown together with the observations of \citet{Panic10}.}
\label{fig:colineshape}
\end{figure}

The predicted shapes of the CO $J=3-2$, $J=10-9$, $J=16-15$ and $J=30-29$ lines are shown in Figure \ref{fig:colineshape}. For CO $J=3-2$, the profile is given for the APEX beam, while the higher-$J$ lines are calculated for the Herschel beam. The HIFI heterodyne spectrometer onboard Herschel (\citealt{deGraauw10}) can spectrally resolve lines down to $<0.1$ km s$^{-1}$ and covers the frequency range of CO $J=5-4$ up to CO $J=16-15$. For illustration purposes, we however also show CO $J=30-29$, which might eventually be accessible to resolved observations with the GREAT instrument onboard SOFIA.

The modeled CO $J=3-2$ line agrees well in terms of width with the spectra observed by \citet{Panic10}. A slight asymmetry in the observed spectra with a stronger blue-shifted peak is however not reproduced by the model. \citet{Panic10} attributed the asymmetry to an asymmetric temperature structure that might be due to a warped inner disk (\citealt{Quillen06}). Modeling these asymmetries is beyond the scope of this study and would require tighter observational constraints on the density structure of the disk.

The CO $J=16-15$ and $J=30-29$ lines are predicted to be considerably broader than those of CO $J=3-2$ and $J=10-9$. While CO $J=3-2$ and $J=10-9$ both have a width of $\sim 5$ km s$^{-1}$, CO $J=16-15$ and $J=30-29$ have widths of $\sim9$ km s$^{-1}$ and $\sim12$ km s$^{-1}$. This can be understood based on the origin of those lines from regions closer to the star. For the assumed Keplerian velocity field around a 2.5 $M_\odot$ star, the projected velocity along the semi-major axis is ${\rm v}_{\rm proj}=\sin(i) \sqrt{M_* G/r} = 31.5 r_{\rm AU}^{-1/2}$ km s$^{-1}$. For a given velocity ${\rm v}_{\rm proj}$ in the observed spectra, $r$ corresponds to the maximum radius from which the line photons can emerge. For the half width of the CO lines discussed here (2.5, 4, and 6 km s$^{-1}$) we get $r=160$, $50$, and $27$ AU, roughly corresponding to the radii of the main emission found in Section \ref{sec:lineorigin}. We note that the CO $J=30-29$ line is broader than expected from the projected Kepler velocity at the inner rim. This is due to thermal broadening of the line, corresponding to more than 1 km s$^{-1}$ for the gas temperatures of a few 1000 K reached in the inner, upper atmosphere. 

We conclude that the width of the CO lines is an interesting indicator of the radial origin of the emission and thus a probe/test of the temperature structure. With current facilities, lines up to CO $J=16-15$ can be spectrally resolved, allowing us to study regions within radii of a few tens of AU as suggested by our results.

%
%
\section{Results: Dependence on parameters} \label{sec:paramstud}

This section presents the results of a parameter study. We vary different input parameters of the model, such as the size of the disk or the amount of gas in the disk in order to understand the dependence of the line fluxes on those parameters. Understanding these dependences will help us to interpret observational trends and also to quantify the robustness of the results, given that large uncertainties enter the modeling (Section \ref{sec:uncerttgas}).  

\subsection{Varying size, gas/dust ratio, and carbon abundance} \label{sec:paramgdrout}

Key parameters for the gas modeling are the total amount of gas and the size of the disk. In this section, we vary the outer radius of the gas disk to be either 200 or 400 AU and assume that gas/dust $=20$ or $100$. A disk outer radius of $r_{\rm out}=400$ AU and gas/dust $= 20$ ($=100$) yields a gas mass of $8 \times 10^{-4}$ $M_{\odot}$ ($4 \times 10^{-3}$ $M_{\odot}$). For an outer radius of $r_{\rm out}=200$ AU, the gas mass is $4 \times 10^{-4}$ $M_{\odot}$ ($2 \times 10^{-3}$ $M_{\odot}$), thus a factor of two lower than for $r_{\rm out}=400$ AU. 

The outer radius of the disk is difficult to determine, as discussed in \citet{Mulders11}. For HD 100546, scattered light has been detected out to $\sim 1000$ AU (\citealt{Ardila07}), but it remains unclear whether this emission originates from the disk or a diffuse remnant envelope. Thus, the disk might be significantly smaller. Modeling the low-$J$ CO lines, \citet{Panic10} constrained the size of the gas disk to be $\sim 400$ AU and the gas mass to be $10^{-3}$ $M_\odot$. \citet{Mulders11} employed an exponential cutoff in surface density outside 350 AU. In other disks, \citet{Hughes08} found that CO $J=3-2$ can be more clearly explained when such a exponential cutoff is used. Here, we use a sharp cutoff at $r_{\rm out}=400$ AU (\citealt{Panic10}), but we also consider a disk with a radius of only 200 AU (\citealt{Pantin00}). 

The gas/dust ratio is assumed to be either an ISM value of 100 or a lower value, as previously suggested by \citet{Chapillon10} to explain the non-detection of [\ion{C}{I}] 370 $\mu$m and 609 $\mu$m lines towards the Herbig Ae star CQ Tau. We adopt gas/dust $=20$, which is similar to their choice. As mentioned before, these gas/dust ratios only refer to the population of small grains and are thus upper limits.

The carbon budget of the gas depends on the elemental depletion on dust grains. The warm ISM shows signs of little or no depletion of oxygen and we thus assume that all oxygen not locked up in silicates is in the gas phase. We note that oxygen may freeze-out as molecules onto the dust grains, e.g. as water ice, but this happens only in the cold shielded regions of the disk (see Section \ref{sec:physstruc}). Carbon, however, may be depleted by a considerable fraction by locking it up in refractory material. We chose a carbon abundance of $\delta_{\rm C} \sim 0.1$, corresponding to an abundance of $2.4 \times 10^{-5}$ relative to hydrogen\footnote{$\delta_{\rm C}$ is defined as the fraction of carbon in the gas-phase (See Section \ref{sec:repmod})}. In this way, we can study the degeneracy between having a higher gas/dust ratio of 100 compared to 20 and a lower carbon gas phase abundance of $\delta_{\rm C} = 0.1$ compared to $\delta_{\rm C} \sim 0.5$. For gas/dust $=100$, we also run models with $\delta_{\rm C} = 0.05$.

\begin{figure*}[htb]
\sidecaption
\includegraphics[width=0.7\hsize]{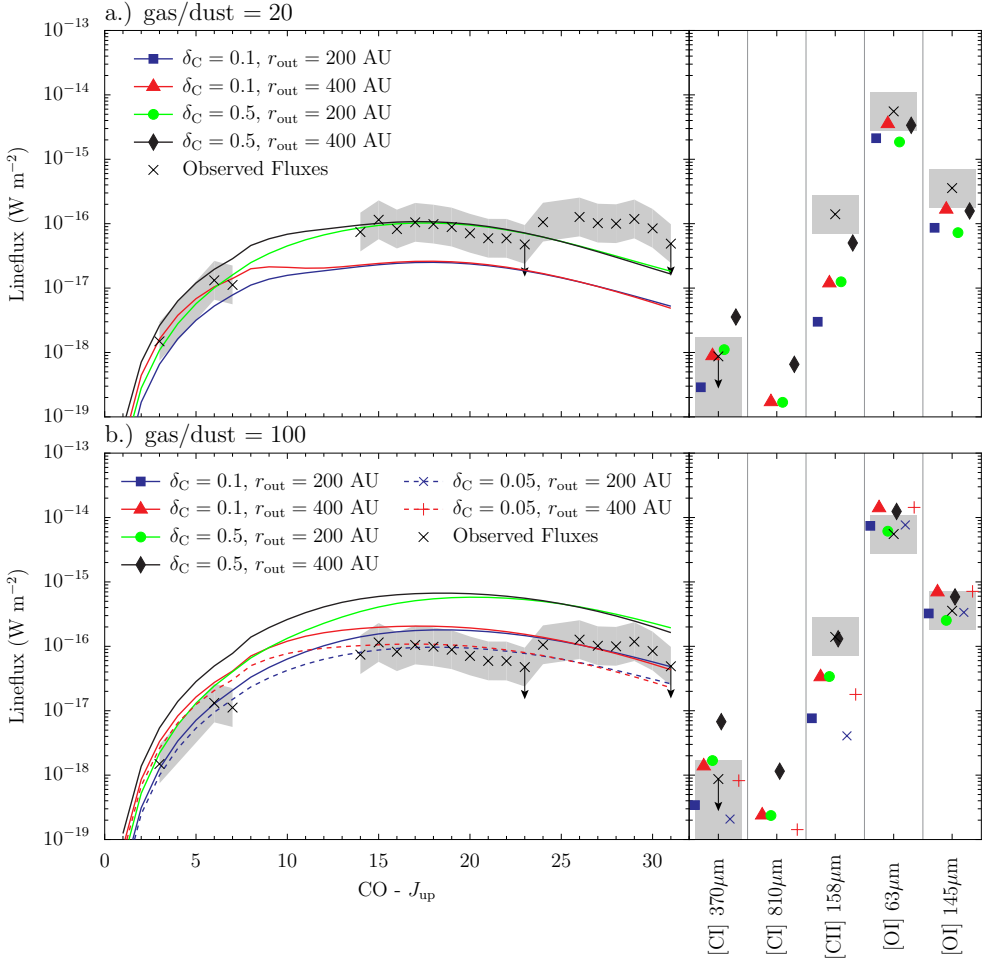}
\caption{Integrated line fluxes for models with different gas/dust ratios, outer radius of the disk, and carbon abundances in the gas phase. \textbf{a.)} With gas/dust $=20$ \textbf{b.)} With gas/dust $=100$.}
\label{fig:comp_gas_carbon_rout}
\end{figure*}

In Figure \ref{fig:comp_gas_carbon_rout}, we show the integrated line fluxes for the models with different radii, gas/dust ratios, and carbon abundances in the gas, as in Figure \ref{fig:fluxreference}. We first discuss the models with gas/dust $=20$ shown in the upper panel of the figure. The low-$J$ and high-$J$ lines of CO have different behaviors. As expected, the high-$J$ lines do not depend on the outer radius of the disk since the radiation emerges from the inner disk. The optically thin high-$J$ lines however scale with the amount of carbon in the gas. We note that this is already true for mid-$J$ lines that are marginally optically thick. The optically thick low-$J$ lines depend less on the amount of carbon, with a change between the model of the same radius of about a factor of two. We note that these optically thick lines may still depend on the abundance, since e.g. the line wings may remain thin. The low-$J$ lines also depend on the outer radius of the disk. Changing the radius from 200 AU to 400 AU increases the area of the disk by about a factor of four. The low-$J$ lines however only change by about a factor of two. This is because a considerable part of the emission originates from within 220 AU (Section \ref{sec:lineorigin}).

As expected, the fine-structure lines of [\ion{O}{I}] for gas/dust $=20$ do not change with the amount of carbon in the gas. They however slightly scale with the outer radius by about the same factor as the low-$J$ lines of CO. This can be understood by these lines emerging from within about the same radial region, as discussed in Section \ref{sec:lineorigin}. 

The [\ion{C}{I}] and [\ion{C}{II}] lines depend on both the outer radius and the amount of carbon in the gas phase by similar amounts. This is the result of both lines being optically thin and emerging from approximately the same region. The line fluxes indeed scale with the amount of carbon in the atmosphere, while they change by about a factor of four with radius.

The models with gas/dust $=100$, shown in the lower panel of Figure \ref{fig:comp_gas_carbon_rout}, exhibit very similar dependences on the outer radius and amount of carbon in the gas phase. How do the lines change with the larger amount of gas in the atmosphere? The optically thin high-$J$ lines of CO approximately scale with the amount of gas. However, since the thermal balance is also affected by the higher density, leading to slightly higher temperatures, the lines increase by more than a factor of five. As expected, the low-$J$ lines scale by a smaller factor, owing to optical depth effects. The [\ion{O}{I}] fluxes approximately scale with gas mass. The carbon fine-structure lines scale by a smaller factor, as the higher density moves the C$^+$/CO transition up in the atmosphere resulting in a smaller increase in the column density. 

The line ratio [\ion{C}{I}]/[\ion{C}{II}] is about a factor of two lower for the higher gas/dust ratio. This is due to different branching ratios in the C$^+$ destruction. The radiative association of C$^+$ with H$_2$ leads to CH$_2^+$, while grain surface reactions of C$^+$ with hydrogenated PAHs (PAH:H) leads to CH$^+$. Following a chain of reactions with H$_2$ and electron recombinations, CH$^+$ and CH$_2^+$ lead to CH and CH$_2$. For higher densities, CH and CH$_2$ are turned into CO rather than being photodissociated (see \citealt{Tielens85}). This only slightly shifts the CO transition, thus affects the CO fluxes only slightly, but does affect the [\ion{C}{I}] lines.

There is some degeneracy between models having more gas (gas/dust $=100$) but less carbon ($\delta_{\rm C}=0.1$) and models with less gas (gas/dust $=20$) and more carbon ($\delta_{\rm C}=0.5$), since they have the same amount of carbon in the atmosphere. These models with the same outer radius agree more closely than about a factor of two in the low-$J$ to mid-$J$ lines of CO and [\ion{C}{II}] (see also Figure \ref{fig:comp_gas_carbon_rout_rv31}, models with $R_{\rm V}=5.5$). The high-$J$ lines of CO exhibit larger changes. As discussed before, neutral carbon changes by about a factor of three.

\subsection{Varying the dust opacity} \label{sec:paramdust}

The dust opacities at UV wavelengths are another key parameter of the gas, as they control the penetration of the high-energy photons into the disk atmosphere. This affects both the chemistry and thermal balance, for example by means of the photodissociation/ionization of molecules or the photoelectric effect on the dust grains. The opacity at short wavelengths is dominated by VSGs and PAHs.

\begin{figure*}[htb]
\sidecaption
\includegraphics[width=0.7\hsize]{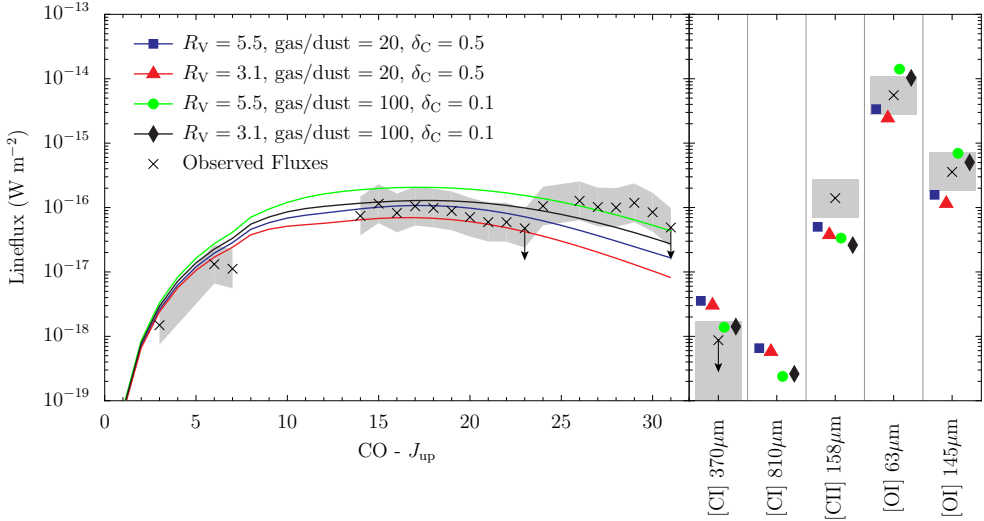}
\caption{Same figure as Figure \ref{fig:comp_gas_carbon_rout}, but either assuming $R_{\rm V}=3.1$ or $R_{\rm V}=5.5$ dust opacities (Figure \ref{fig:dustopac}). The outer radius of the models is 400 AU.}
\label{fig:comp_gas_carbon_rout_rv31}
\end{figure*}

\begin{figure*}[htb]
\sidecaption
\includegraphics[width=0.7\hsize]{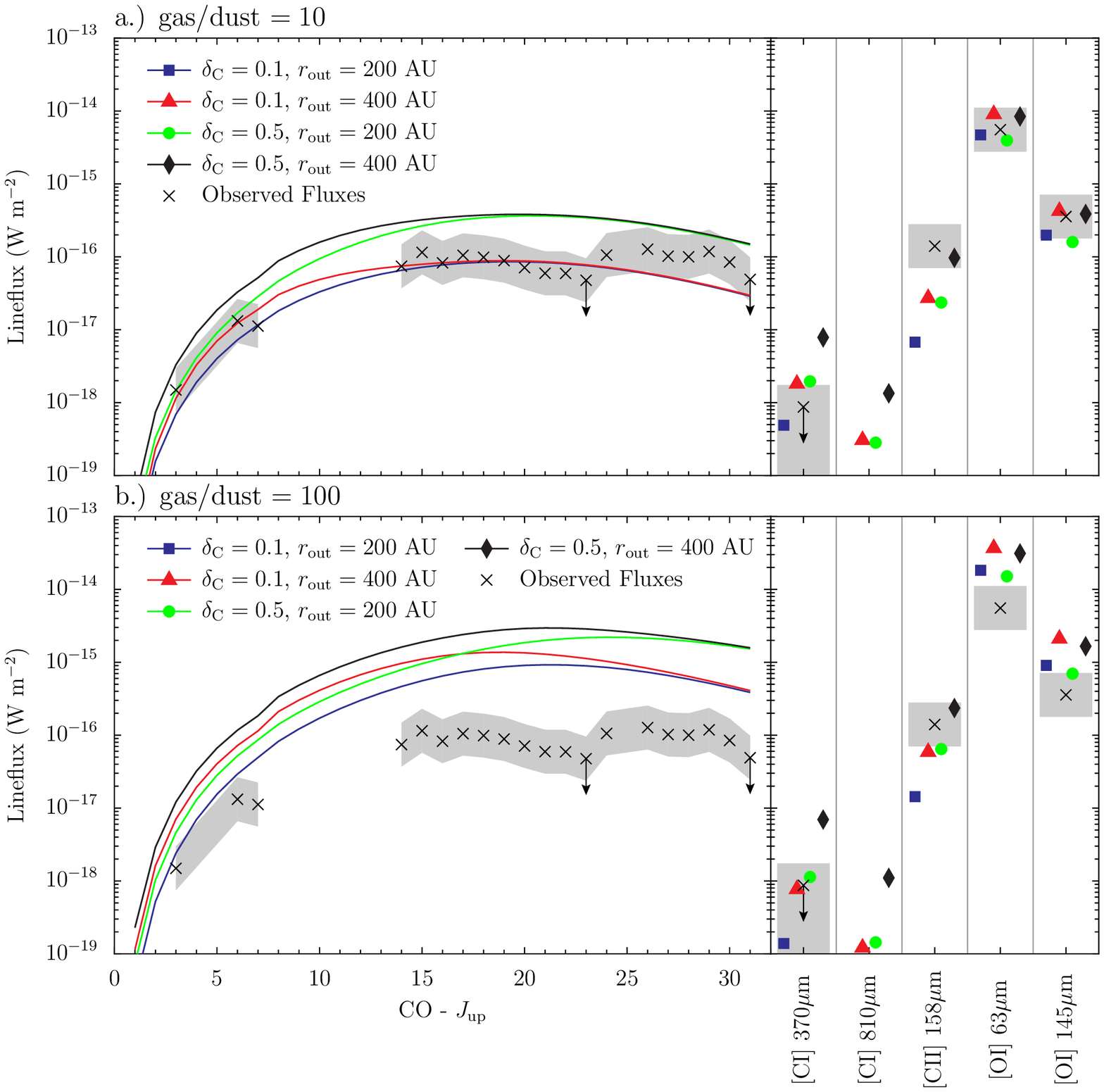}
\caption{Same figure as Figure \ref{fig:comp_gas_carbon_rout}, but assuming 0.1-1.5 $\mu$m dust grains (\citealt{Mulders11}). \textbf{a.)} With gas/dust $=10$ \textbf{b.)} With gas/dust $=100$.}
\label{fig:comp_gas_carbon_rout_mulders}
\end{figure*}

Observationally, the size of the smallest grains is difficult to determine. For HD 100546, indications of grain processing and larger minimum grain sizes were found by \citet{Ardila07}. \citet{Bouwman03} suggested that the size distributions vary with radii, employing a population of smaller grains in the inner 40 AU and larger grains in the outer disk. Using such a grain distribution, \citet{Benisty10} fit the SED using silicate dust grains and a surface layer of $0.05-1$ $\mu$m grains at $13-50$ AU and an outer disk consisting of $1$ $\mu$m - $1$ cm size grains. In contrast \citet{Mulders11} manage to reproduce the SED by assuming a single grain distribution of $0.1-1.5$ $\mu$m size, but using a vertical structure in hydrostatic equilibrium rather than a fixed scale height. In addition to radially dependent dust properties, dust settling may also introduce a vertical stratification of dust properties (e.g. \citealt{dAlessio06}), having a similar effect on the gas as a larger scale height of the gas structure and thus a higher gas/dust ratio in the upper atmosphere (\citealt{Jonkheid06}). 

In this section, we study the influence of the dust opacity at UV wavelengths on the line fluxes. A detailed study of settling and radially dependent dust properties (e.g. \citealt{Birnstiel10}) is beyond the scope of this study and we only globally change the dust properties at UV/Visual wavelengths. We assume the dust opacity laws given in Section \ref{sec:modhd100} (Figure \ref{fig:dustopac}). These are either ISM-like grains with $R_{\rm V}=3.1$, corresponding to a minimum grain size of $\sim 50$ \AA, grains with $R_{\rm V}=5.5$ corresponding to some coagulation and grain growth and considerably larger grains with a minimum size of 0.1 $\mu$m. The models with a minimum size of 0.1 $\mu$m do not contain any VSGs/PAHs and the opacity at UV/visual wavelengths is thus considerably smaller. Photodestruction of VSGs to PAHs in the upper atmosphere (\citealt{Berne09}) leads to some additional vertical change in the UV opacity, even though their mass extinction coefficients in the UV are similar (e.g. \citealt{Pontoppidan07}, Figure 2). We thus consider this section as an approach to studying the influence of different UV penetration lengths on the chemistry, independent of whether the absorption is due to VSGs or PAHs or the absence of both.

In Figure \ref{fig:comp_gas_carbon_rout_rv31}, we show the line fluxes obtained using the $R_{\rm V}=3.1$ and $R_{\rm V}=5.5$ dust opacities. The outer radius of the models is 400 AU, and we adopt either gas/dust $=100$ and $\delta_{\rm C}=0.1$ or gas/dust $=20$ and $\delta_{\rm C}=0.5$. All fine-structure lines change by less than 50 \% by using the two different opacity laws. The low-$J$ lines of CO change similarly little, but the high-$J$ lines deviate by up to about a factor of two. The changes are very similar to those for all other combinations of outer radii, gas/dust ratio, and carbon abundance.

\begin{figure*}[htb]
\includegraphics[width=1.0\hsize]{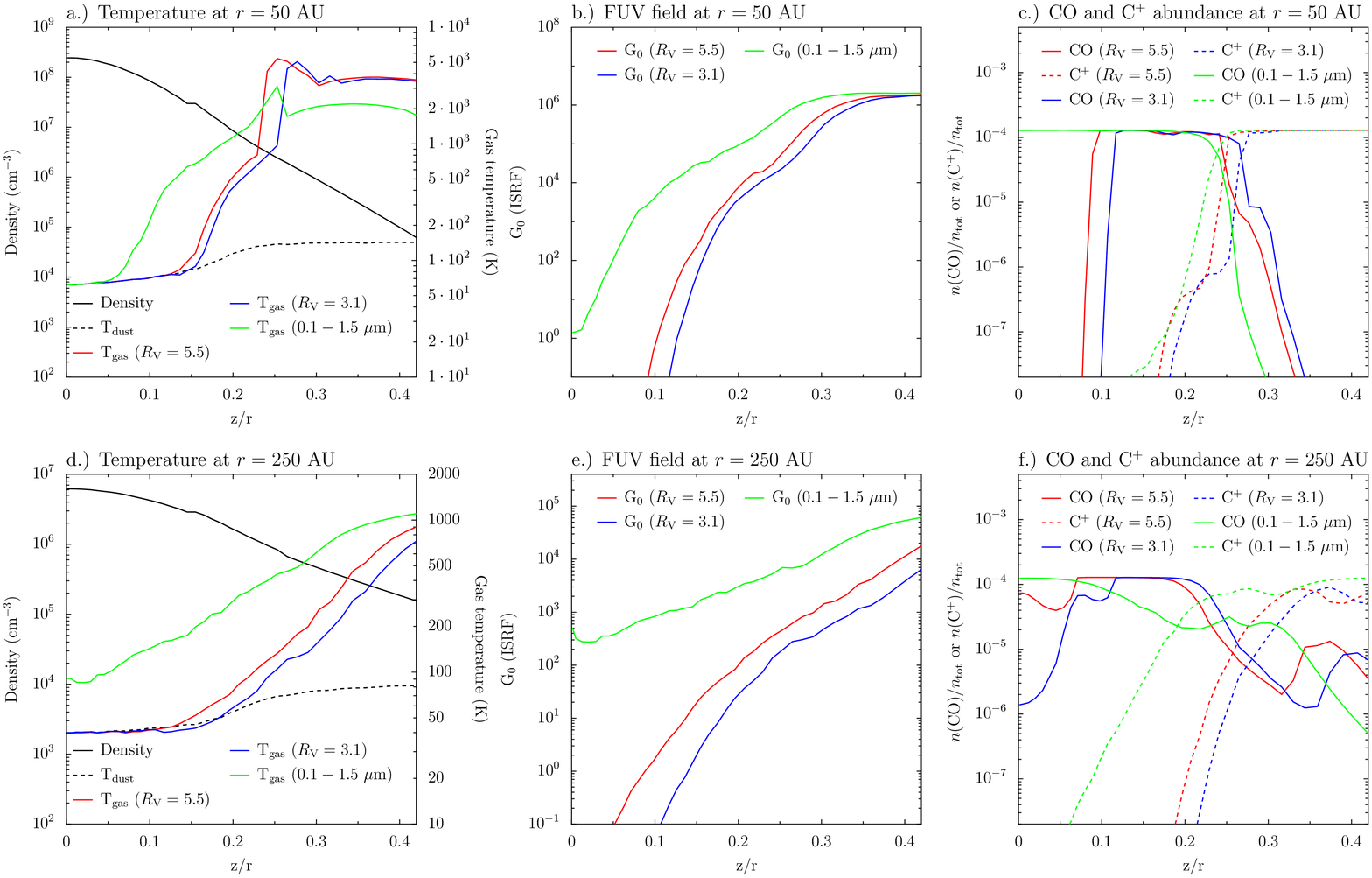}
\caption{Vertical cuts through the disk at a radius of 50 AU (top panels) and 250 AU (lower panels). Green lines show values obtained using the 0.1-1.5 $\mu$m grains, blue and red lines indicate models calculated with $R_{\rm V}=5.5$ and $R_{\rm V}=3.1$, respectively. \textbf{a.)/d.)} Gas density, gas an dust temperature \textbf{b.)/e.)} FUV field in units of the interstellar radiation field \textbf{c.)/f.)} Abundance of CO and C$^+$ relative to the hydrogen density.}
\label{fig:cut_50250}
\end{figure*}

The results obtained using the much larger 0.1-1.5 $\mu$m grains are given in Figure \ref{fig:comp_gas_carbon_rout_mulders}. For gas/dust $=100$, the line fluxes are much larger than the models with $R_{\rm V}=5.5$ dust opacities, with increases of up to an order of magnitude. Thus, the observed line fluxes can no longer be reproduced, not even with the lower carbon abundance. To roughly fit the observations, we thus show models with gas/dust $=10$ in a second panel. The dependences of the models with large grains on the outer radius and carbon abundance is similar those of to the $R_{\rm V}=5.5$ models. 

We conclude that the dust opacity, particularly that of the small grains governing the penetration of the UV radiation into the disk atmosphere, has major implications for the line fluxes. What is the reason for this strong dependence? In Figure \ref{fig:cut_50250}, we show vertical cuts through the disk in density, temperature, FUV field, and fractional abundance of CO and C$^+$ for the models with different dust opacity laws. Comparing the results obtained with $R_{\rm V}=3.1$ and $R_{\rm V}=5.5$ shows that while the gas temperatures in the upper atmosphere are similar, the transition to the cold atmosphere, where gas and dust temperature are coupled, occurs at greater depth. Similarly, the C$^+$/CO transition occurs at greater depths within the disk. Both results are caused by the FUV field that can penetrate deeper into the disk. While the temperature and abundance structures are qualitatively similar for the $R_{\rm V}=3.1$ and $R_{\rm V}=5.5$ models, they are considerably different for the larger grains: The gas temperature is still coupled to the dust temperature in the inner midplane (50 AU). Further out (at 250 AU), the gas temperature however no longer couples to the dust temperature. This is consistent with the much larger UV field in the midplane, which also explains the shift of the C$^+$/CO transition deeper into the disk. Thus, larger grains allow UV radiation to penetrate more deeply into the disk to regions of higher density and also to heat the gas to higher temperatures. The combination of both leads to the considerable increase in line flux.

\subsection{Varying PAH abundance} \label{sec:parampah}

Given the uncertain abundance of PAHs and the difficulties in simultaneously reproducing the [\ion{C}{II}] fluxes and [\ion{C}{I}] upper limits, we explore the effect of the PAH abundance on the line fluxes in this section. The PAH abundance affects the models in two ways. First, the ionization structure is changed by PAHs by means of recombination reactions such as C$^+$ + PAH$^0$ $\rightarrow$ C + PAH$^+$, with PAH$^0$ being neutral and PAH$^+$ positively charged PAHs. Second, the gas heating depends on the PAHs by means of the photoelectric effect on small PAHs.

\begin{figure*}[htb]
\sidecaption
\includegraphics[width=0.7\hsize]{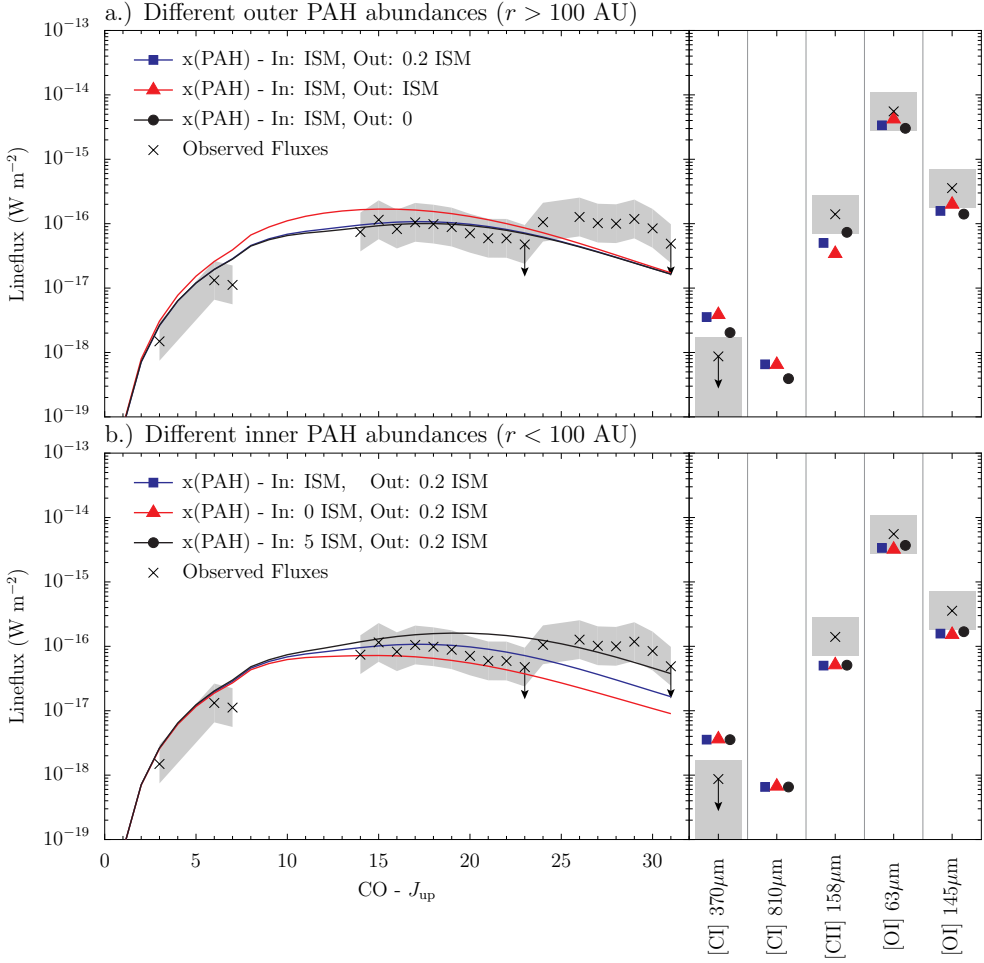}
\caption{Dependence of the integrated line fluxes on the PAH abundance. \textbf{a.)} The abundance of PAH outside 100 AU is varied. \textbf{b.)} The PAH abundance within 100 AU is varied.}
\label{fig:pahvary}
\end{figure*}

The emission of PAH can be directly observed by its bands in the infrared (e.g. \citealt{Geers07a}). This emission however only traces regions where PAHs are excited by UV photons. An abundance and excitation study by \citet{Visser07} shows that in disks around Herbig Ae stars, most of the emission from intermediate-size PAHs emerge from the uppermost layer, above the $\tau_{\rm V}=1$ surface. Thus, PAHs do not trace down to regions where the C$^+$/C transition is affected by the PAH charge exchange. The PAH abundance is thus rather uncertain.

Figure \ref{fig:pahvary} presents line fluxes obtained with models that contain different PAH abundances in the outer ($r>100$ AU) and inner ($r<100$ AU) part of the disk. The outer radius of the disk is 400 AU, an ISM carbon abundance, and gas/dust $=20$, corresponding to the representative model presented in Section \ref{sec:repmod}.

Changing the PAH abundance at radii $> 100$ AU from 20 \% ISM to no PAHs, hardly changes the CO ladder. In addition, the [\ion{O}{I}] fine-structure lines are hardly affected. The [\ion{C}{II}] line however slightly increases and [\ion{C}{I}] decreases by a factor of about two, owing to the missing charge exchange converting C$^+$ to C. Changing the PAH abundance in the outer disk to the ISM abundance increases the mid-$J$ lines of CO, owing to the greater heating caused by the higher abundance of PAHs. We note that the high-$J$ lines of CO are less affected than the lower-$J$ lines, as they mostly emerge at radii $< 100$ AU. The fine-structure lines are only slightly affected with [\ion{C}{II}] being slightly weaker and [\ion{C}{I}] slightly stronger, owing to the greater charge exchange. We also tested the influence of the charge exchange with sulfur (C$^+$ + S $\rightarrow$ C + S$^+$), which had been traditionally found to influence [\ion{C}{I}] in PDRs (\citealt{Hollenbach97}, Section 4.4). For the dense PDR of the disk atmosphere including a density gradient, the influence of sulfur charge exchange is however found to be smaller than that for PAHs.

Varying the PAH abundance in the inner 100 AU does not affect the atomic fine-structure lines. The high-$J$ lines of CO however change by about a factor of two, if the PAH abundance is higher by a factor of five or set to zero. This is a result of either the higher or lower gas temperature caused by the contribution of photoelectric heating on PAHs to the thermal balance. 

\subsection{The stellar input spectra} \label{sec:stespec}

\begin{figure*}[htb]
\sidecaption
\includegraphics[width=0.7\hsize]{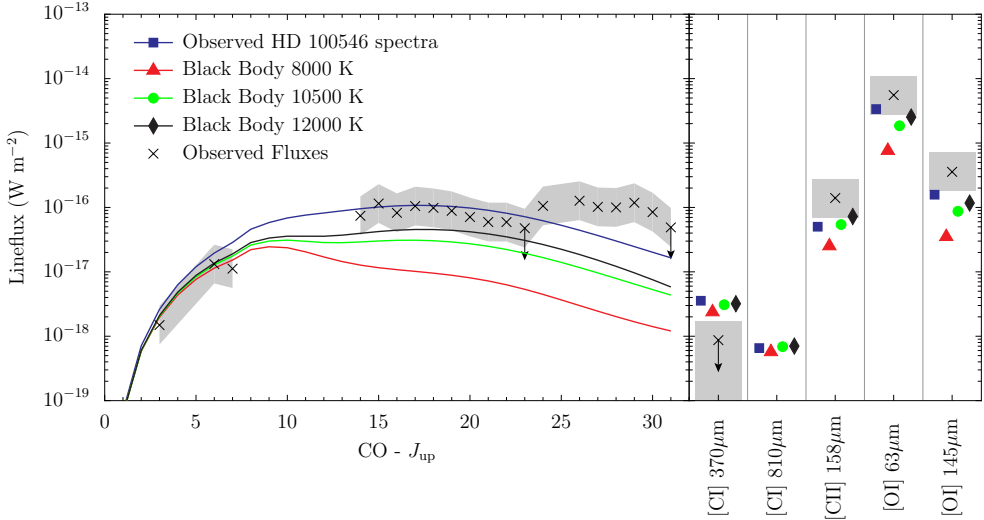}
\caption{Line fluxes of models with different stellar radiation fields.}
\label{fig:comp_spec}
\end{figure*}

The stellar UV radiation field is an important input parameter for the model, as both the heating and the chemistry depend on the input radiation field. We use observed and dereddened IUE/FUSE spectra. However, the dereddening is somewhat uncertain, particularly for shorter wavelengths. We thus also run models using a blackbody of 8000 K, 10500 K ($T_{\rm eff}$ of HD 100546), and 12000 K (Figure \ref{fig:ste_uv_spec}). The blackbody was scaled to preserve the bolometric luminosity. Thus, the luminosity in the FUV band changed. While $L_{\rm FUV}/L_{\rm bol} \sim 0.27$ for the observed FUV spectra, it is $L_{\rm FUV}/L_{\rm bol} \sim 0.024$ (0.095, 0.15) for the blackbody of 8000 (10500, 12000) K.

In Figure \ref{fig:comp_spec}, the line fluxes of the models with the different input spectra are shown. Models with a lower FUV luminosity $L_{\rm FUV}$ have less heating and thus the mid/high-$J$ CO and [\ion{O}{I}] lines are weaker (Section \ref{sec:warmatmos}). The [\ion{C}{I}]/[\ion{C}{II}] ratio however changes only slightly with the UV color. We note that both C and CO are photoionized/photodissociated at similar wavelengths ($\lambda \lesssim $1100 \AA) and their rates thus depend similarly on the UV color. For much colder UV spectra with very few CO dissociating photons, the gas temperature is lower owing to more efficient cooling by CO than for the atomic fine-structure lines. We conclude that for the lines discussed in this work, the UV spectrum affects the lines mostly by means of the photoelectric heating on PAHs/VSGs, which is less efficient for lower $L_{\rm FUV}$. The [\ion{C}{I}]/[\ion{C}{II}] ratio remains however similar and the results related to their flux are unaffected by an uncertainties in the dereddening of the UV spectra.

\subsection{Dependence on the H$_2$ formation rate} \label{sec:deph2}

\begin{figure*}[htb]
\sidecaption
\includegraphics[width=0.7\hsize]{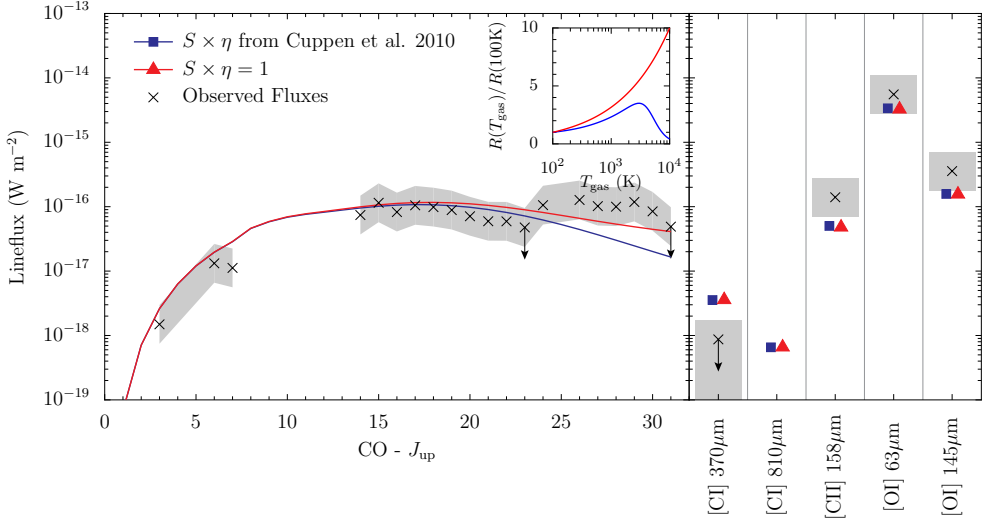}
\caption{Fluxes for models assuming different sticking coefficients/formation efficiencies to calculate the H$_2$ formation rates. The temperature dependence of the rate is given in the inset.}
\label{fig:diffh2}
\end{figure*}

A major uncertainty in the chemistry is the H$_2$ formation rate at high temperature. This rate may affect the line fluxes both in terms of the change in the H/H$_2$ transition, which also affects the C$^+$/C/CO transition and the thermal balance. The heating and cooling processes affected by H$_2$ are cooling by H$_2$ lines and heating by means of UV pumping followed by collisional deexcitation, formation, and photodissociation heating. Uncertainties enter the H$_2$ formation rate in terms of the minimum grain size that determines the grain surface on which H$_2$ may form, but also through the sticking coefficient and the formation efficiency. The sticking coefficient gives the probability that H atoms stick onto the surface rather than bounce back or evaporate before reacting to H$_2$. This sticking coefficient and the formation efficiency depend on various parameters (e.g. \citealt{Hollenbach99}). It may either decrease with the gas temperature when H$_2$ formation on physisorbed sites dominates, or stay constant when at high gas temperature the neighboring chemisorbed H atoms react and leave the surface relatively bare. 

In this section, we calculate the representative model with a sticking coefficient times formation efficiency $S\times \eta=1$, that is independent of temperature. We then compare the result to our ``standard'' H$_2$ formation rate using the $S \times \eta$ reported by \citet{Cuppen10}, in order to capture the theoretical work of \citet{Cazaux04}, \citet{Cuppen05} as well as the observational evidence from \citet{Habart04}. Figure \ref{fig:diffh2} shows the line fluxes calculated with the models assuming different formation rates. The inset in that figure gives the temperature dependence of the H$_2$ formation rate.

Neither the low-$J$ to mid-$J$ CO lines nor the atomic fine-structure lines depend on the H$_2$ formation rate. The high-$J$ lines of CO, however, increase by a factor of $\sim 3$ when the sticking coefficient is assumed to be independent of gas temperature. This is consistent with in the inner disk ($r \lesssim 70$ AU, $z/r \gtrsim 0.15$) hydrogen being fully molecular (see Figure \ref{fig:physstruct} and \ref{fig:chemstruct}), which increases both the gas temperature and CO abundance in this region. We conclude that the predictions of the CO ladder are robust for the low-$J$ and mid-$J$ lines, although some differences due to the uncertain H$_2$ formation rate are expected for the high-$J$ lines with $J_{\rm up} \gtrsim 25$.

%
%
\section{Discussion} \label{sec:discuss}

\subsection{Comparison with observations}

How do the results of the models compare to the observed fluxes? Studying the outer radius, gas/dust ratio, and carbon abundance in Section \ref{sec:paramgdrout} (Figure \ref{fig:comp_gas_carbon_rout}), we find that the mid-$J$/high-$J$ CO lines can be reproduced by either a higher gas/dust ratio and a lower $\delta_{\rm C}$ or a lower gas/dust ratio and a higher $\delta_{\rm C}$. The low-$J$ CO and [\ion{O}{I}] lines emerge from the outer part of the disk, hence scale with the outer radius of the disk. It is thus possible to reproduce these lines for both higher gas/dust and lower gas/dust ratios, if the outer radius is adjusted accordingly. For a gas/dust ratio of 20 and $\delta_{\rm C}=0.5$, a 400 AU disk better reproduces the [\ion{O}{I}] fine-structure lines as the 200 AU disk which underproduces the flux. For a gas/dust ratio of 100 and $\delta_{\rm C}=0.05/0.1$, a smaller disk better reproduces the [\ion{O}{I}] observations, but a 400 AU disk does not considerably overproduce the emission. 

For both gas/dust ratios and outer radii, the modeled low-$J$ lines agree reasonably well with the observations. Independent of the outer radius, the [\ion{C}{II}] line is underpredicted for the models that fit the CO ladder and [\ion{O}{I}]. For gas/dust $=20$, $\delta_{\rm C}=0.5$, and $r_{\rm out}=200$ AU, the [\ion{C}{II}] is only underproduced by about a factor of three, but that model overpredicts the [\ion{C}{I}] line by about a factor of four. However, for gas/dust $=100$ and $\delta_{\rm C}=0.1$, both models with $r_{\rm out}=200$ AU and 400 AU underproduce [\ion{C}{II}], but are within the upper limit to [\ion{C}{I}]. The same is true for the model with $\delta_{\rm C}=0.05$, which however more closely reproduces the CO ladder. Thus, none of the models are able to simultaneously reproduce both the [\ion{C}{II}] line and the upper limit to [\ion{C}{I}]. We conclude that only models with a higher gas/dust ratio and low $\delta_{\rm C} \lesssim 0.1$, are able to simultaneously reproduce the CO ladder, [\ion{O}{I}],  and the upper limit to [\ion{C}{I}] but underpredict [\ion{C}{II}] by a factor depending on the outer radius of the disk.

Changing the dust opacities from $R_{\rm V}=3.1$ to $R_{\rm V}=5.5$ does not change these conclusions (Section \ref{sec:paramdust}, Figure \ref{fig:comp_gas_carbon_rout_rv31}). For models with 0.1-1.5 $\mu$m grains and a much lower UV/visual extinction, the models with gas/dust $=100$ overproduce both the CO ladder and [\ion{O}{I}]. For gas/dust $=10$ and $\delta_{\rm C}=0.1$, both the CO ladder and the [\ion{O}{I}] lines are fit reasonably. The [\ion{C}{II}] line is underproduced, but [\ion{C}{I}] is within the upper limit for both radii. This is similar to our results for the models with a $R_{\rm V}=5.5$ dust opacity law, gas/dust $=100$, and $\delta_{\rm C}=0.05/0.1$. 

Decreasing the PAH abundance in the outer disk (radii $> 100$ AU) mostly affects the [\ion{C}{I}]/[\ion{C}{II}] ratio and a model without PAHs in this region starts to approximate both lines (Section \ref{sec:parampah}, Figure \ref{fig:pahvary}). A higher PAH abundance in the inner disk ($<100$ AU) increases the flux of the mid-$J$ and high-$J$ CO lines and yields a slightly closer agreement with the high-$J$ lines, despite slightly overproducing some mid-$J$ lines. Given the large uncertainties in the gas temperature (Section \ref{sec:uncerttgas}), we however conclude that these changes are not significant. 

Using an H$_2$ formation rate with a sticking coefficient that is independent of temperature slightly increases the high-$J$ lines of CO, which are in closer agreement than using our ``standard'' sticking coefficient that decreases above a certain temperature. However, as discussed before, the uncertain gas temperature most affects the high-$J$ lines and we thus do not consider the change in the high-$J$ lines with H$_2$ formation rate as significant. 

We conclude that the CO ladder together with [\ion{O}{I}] and the upper limit to [\ion{C}{I}] can be reproduced by models with a high gas/dust ratio and low $\delta_{\rm C}$. These models however underproduce [\ion{C}{II}]. Models without PAHs in the outer disk, improve the fit to [\ion{C}{I}] and [\ion{C}{II}] but there is still a factor of two discrepancy for both lines.

\subsection{The [\ion{C}{I}]/[\ion{C}{II}] ratio and the gas carbon abundance} \label{sec:cIcIIratio}

\begin{figure*}[htb]
\sidecaption
\includegraphics[width=0.75\hsize]{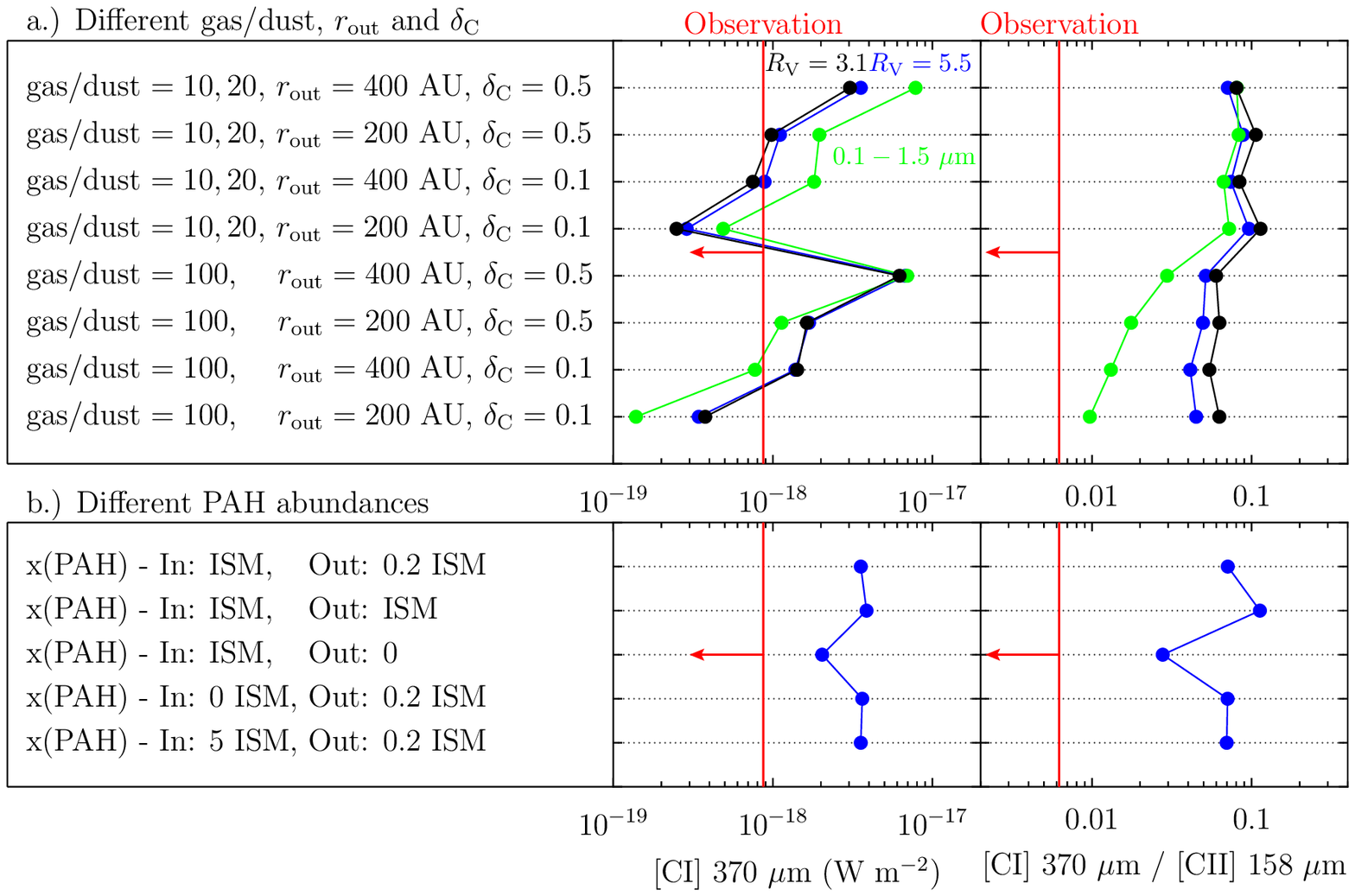}
\caption{Modeled flux of the [\ion{C}{I}] 370 $\mu$m line and the line ratio of [\ion{C}{I}] 370 $\mu$m / [\ion{C}{II}] 158 $\mu$m. Note that in panel \textbf{a.)}, gas/dust $=10$ is given for the $0.1-1.5$ $\mu$m opacities, and gas/dust $=20$ for $R_{\rm V}=3.1,5.5$. The symbol $\delta_{\rm C}$ is the fraction of carbon in the gas-phase and $r_{\rm out}$ is the outer radius of the disk. In panel \textbf{b.)}, x(PAH) is the PAH abundance relative to the ISM abundance within and outside 100 AU.} 
\label{fig:flux_cratio}
\end{figure*}

Given the difficulties in simultaneously reproducing the [\ion{C}{II}] line and the [\ion{C}{I}] upper limit, it is interesting to study the dependence of the two lines on the parameters independent of the other lines. The modeled fluxes of the [\ion{C}{I}] 370 $\mu$m line and the ratio to the [\ion{C}{I}] 158 $\mu$m line are shown in Figure \ref{fig:flux_cratio}, which summarizes Figures \ref{fig:comp_gas_carbon_rout}, \ref{fig:comp_gas_carbon_rout_mulders} and \ref{fig:pahvary}.

The dependence on $r_{\rm out}$, gas/dust, and $\delta_{\rm C}$ is shown in the upper panel of Figure \ref{fig:flux_cratio} for models with the different dust opacities used in this work. The [\ion{C}{I}] 370 $\mu$m line and the [\ion{C}{I}] 370 $\mu$m / [\ion{C}{II}] 158 $\mu$m ratio depend similarly on $r_{\rm out}$, gas/dust and $\delta_{\rm C}$ for the different opacity laws. Exceptions are the large grain models owing to the UV photons penetrating further into the disk, thus dissociating precursors of CO, such as CH and CH$_2$, and affecting C/C$^+$. The [\ion{C}{I}] 370 $\mu$m line varies by about a factor of four between the models with outer radii of 200 AU and 400 AU and a similar factor between $\delta_{\rm C}=0.1$ and 0.5. The [\ion{C}{I}] 370 $\mu$m /[\ion{C}{II}] 158 $\mu$m line ratios are almost independent of both the outer radius and carbon abundance, but depend slightly on the gas/dust ratio, again since CH and CH$_2$ are more likely to be converted into CO than photodissociated at higher density, thus decreasing the [\ion{C}{I}] fluxes. Thus, by increasing the gas mass by a factor of five from gas/dust of 20 to 100 only increases the [\ion{C}{I}] by about a factor of two. This reduces the [\ion{C}{I}] 370 $\mu$m /[\ion{C}{II}] 158 $\mu$m line ratio by about a factor of two. For the larger grains, this effect becomes much stronger because the lower UV extinction moves the C$^+$/C transition deeper into the disk, to higher densities. We note however, that the models with large grains and gas/dust $=100$ overproduce CO and [\ion{O}{I}] (Section \ref{sec:paramdust}).

The PAH abundance within 100 AU does not affect the [\ion{C}{I}] and [\ion{C}{II}] fluxes. However, if PAHs are absent in the outer part of the disk ($> 100$ AU), the [\ion{C}{I}] fluxes decrease by about a factor of two, and the [\ion{C}{I}] 370 $\mu$m /[\ion{C}{II}] 158 $\mu$m line ratio decreases by about a factor of three. As discussed before, this is due to the lower charge exchange from C$^+$ to C by PAHs. Similarly, increasing the the PAH abundance from 20 \% ISM to ISM abundances leads to a stronger charge exchange and thus a larger line ratio.

In spite of all these parameter changes, the observed [\ion{C}{II}] 158 $\mu$m line and the observed line ratio [\ion{C}{I}] 370 $\mu$m / [\ion{C}{II}] 158 $\mu$m cannot be reproduced by our models. Our closest match is found for the model without PAHs in the outer part of the disk (Figure \ref{fig:pahvary}), but that model still differs by a factor of about two for both lines. The [\ion{C}{I}] 370 $\mu$m / [\ion{C}{II}] 158 $\mu$m line ratio is surprisingly independent of the parameters, which can be understood by a similar region of origin of the two lines found in Section \ref{sec:lineorigin}.

\subsection{The origin of the [\ion{C}{II}] emission?} \label{sec:cIIemission}

Does the observed [\ion{C}{II}] 158 $\mu$m emission really originate from the disk? Since C$^+$ is the main state of carbon in diffuse clouds and owing to the low critical density of the 158 $\mu$m line, how much could a tenuous remnant envelope or a diffuse foreground cloud contribute to the emission? We note that the discussion here assumes that the emission is centered on the source, within a $\sim 10''$ (1000 AU) diameter region, since extended emission has already been subtracted (Section \ref{sec:observ}). 

If the extinction towards the star results from material only seen in the central pixel, how much could that contribute to the [\ion{C}{II}] emission? The maximum amount of emission to be expected from the foreground can be calculated from the foreground extinction of $A_{\rm V}=0.36$ mag (\citealt{Ardila07}). Following Appendix \ref{sec:app_cplus}, we obtained a flux of $\sim 2.5 \times 10^{-16}$ W m$^{-2}$ in the Herschel beam, using a conversion of $1.9\times10^{21}$ cm$^{-2}$ $A_{\rm V}^{-1}$ (\citealt{Bohlin78}), a carbon fractional abundance of about $1.3 \times 10^{-4}$, and a gas temperature in the foreground above the upper level energy of the [\ion{C}{II}] 158 $\mu$m line ($91$ K). This would already exceed the flux reported in Table \ref{tab:flux} by 60 \%.

How much emission could come from a compact diffuse remnant envelope? Scattered light is detected out to a distance of 1000 AU (\citealt{Ardila07}) and could belong to either the disk or a remnant envelope. If a tenuous remnant envelope is exposed sufficiently to UV radiation, most of its carbon is in the form of C$^+$. This nebula cannot contain CO, since this would fill in the central emission of the observed double-peak line shape (\citealt{Panic10}). While the density structure of such a remnant envelope is not known from observations, assuming a ring size in the range $\sim 400-1000$ AU at a density of $\sim 10^4$ cm$^{-3}$ and a thickness of $\sim 800$ AU would yield a flux of up to $1.4 \times 10^{-17}$ W m$^{-2}$. Thus, both a foreground cloud and a remnant envelope might well contribute considerably to the [\ion{C}{II}] emission. We note that no [\ion{C}{I}] emission is expected from this component, since for typical columns of C towards diffuse clouds ($\lesssim 3 \times 10^{15}$ cm$^{-2}$, \citealt{Jenkins79}), the [\ion{C}{I}] 370 $\mu$m line is below the detection limit of APEX.

This rough estimate shows that both foreground and remnant envelope could add substantially to the [\ion{C}{II}] emission, making the association of the observed flux with the disk not obvious. This question can however only be addressed with velocity resolved spectra of [\ion{C}{II}] 158 $\mu$m, obtained with the HIFI instrument onboard Herschel.

\subsection{The carbon budget of the disk atmosphere}Ê\label{sec:carbbudget}

Our results indicate that for normal gas/dust ratios, the carbon abundance is low at $\delta_{\rm C} \lesssim 0.1$. How does the gas-phase carbon abundance change as the gas evolves from diffuse clouds to dense clouds, collapsing envelopes,  and eventually disks? At the earliest stage, in diffuse clouds, the carbon gas-phase abundance can be measured directly by the UV absorption of \ion{C}{II}, without referring to a dust model (e.g. \citealt{Sofia04,Sofia11}). The measured amount of volatile carbon, which is not bound in dust, is constrained to be between $\sim$ 40 \% and 60 \% ($\delta_C=0.4$ to $0.6$), with some differences between the measurements of the strong 1334 \AA\, feature (\citealt{Sofia11}) and the weak intersystem feature at 2325 \AA\,(\citealt{Sofia04}). Assuming that the refractory material does not change as the gas evolves to cores and finally disks, these micron-sized carbon grains quickly settle down to the midplane of the disk. Thus, if a process removes gas from the upper disk, that population of carbon dust is not removed from the disk. 

The volatile carbon however undergoes significant changes in the course of this evolution: from diffuse to dense clouds, the main reservoir becomes gaseous CO rather than C$^+$. In pre-stellar cores, it becomes CO ice rather than CO gas (\citealt{Bergin07}). From pre-stellar cores to disks, it either stays in the form of CO ice, adsorbs or desorbs from grains multiple times, or is transformed into other species such as CO$_2$ or CH$_3$OH (\citealt{Visser09a, Visser11}). If gas is removed from the disk by some process, leading to a smaller gas/dust ratio, the volatile carbon is likely to be removed at a rate that is proportional to the quantity of gas. Thus, within the gas, the amount of volatile carbon should remain the same. Inferring a value of $\delta_C$ that is much lower than 0.4 to 0.6 as we find here means that a fraction of volatile carbon has been turned into non-volatile material in the course of the evolution from diffuse clouds to disks. Towards hot cores, where CO ice is supposed to evaporate from dust grains, \cite{AlonsoAlbi10} and \cite{Yildiz10} found evidence of low CO hot core abundances, suggesting some transformation of CO on the surfaces of the grains into other volatile or non-volatile carbon species. For the warm disk discussed here, CO ice cannot be the main carbon reservoir. In addition, no strong CH$_3$OH emission is in general observed in disks, ruling out this possible carbon reservoir. However, some ``mild'' photochemistry could lead to more complex, less volatile organics (\citealt{Oberg09b}). Some of this less volatile organic material could become similar to the so-called ``CHON particles'' detected in comets (\citealt{Kissel86}) after further processing in the disk.

The upper layers of the disk atmosphere, probed in this work, more likely have gas/dust $>100$ rather than $<100$, since settling increases the gas/dust ratio in that region. In addition, as we keep the dust density structure in our work fixed, a small gas/dust ratio would mean that the relative fraction of carbon locked into dust (VSGs/PAHs) increases for small gas/dust ratios, which actually implies that the total carbon abundance is higher than the solar abundance. These arguments drive us to the conclusion that the warm gas in the HD 100546 atmosphere more likely consists of a large gas/dust ratio with a considerably smaller abundance of volatile carbon than in the diffuse ISM.

\subsection{Uncertainties in $T_{\rm gas}$} \label{sec:uncerttgas}

An important caveat of the models presented here is the uncertainty in the derived gas temperature. On the surface of the disk where the gas and dust temperature decouple, the equilibrium of different heating and cooling processes govern the gas temperature. As noted before, the thermal balance is a coupled problem together with the chemistry as line cooling (in [\ion{C}{II}], [\ion{O}{I}], CO, \ldots) depends on the abundances. From PDR modeling, the thermal balance problem is known to be very sensitive to the assumptions made, such as the method used for the line radiative transfer or the H$_2$ physics and chemistry implemented. \citet{Roellig07} compared the gas temperature derived by different PDR models assuming the same chemical network. They found a relatively large scatter, in particular for the model with a high density ($n=10^{5.5}$ cm$^{-3}$) and a strong UV field ($\chi=10^5$) corresponding approximately to the upper atmosphere of the outer disk. The scatter between the models is about a factor of three in $T_{\rm gas}$ (Figure \ref{fig:benchmark_temp}) and similar for the calculated fine-structure line fluxes. In our present study, we found the high-$J$ lines of CO to be more sensitive to the temperature (Section \ref{sec:warmatmos}) and we thus calculated the CO ladder from the CO abundance and gas temperature obtained from the different PDR codes. The scatter in those line fluxes was found to be about 1.5 dex when outliers were removed. The chemical network, which has been taken to be the same than the one used in the benchmark study by \citet{Roellig07}, also introduces uncertainties. \citet{Vasyunin08} performed a Monte Carlo sensitivity analysis and determined the uncertainty in the CO and C$^+$ column density through a disk to be a factor of two or less. 

For our work, we thus consider an agreement of a factor of two between observations and model as a good agreement. Given that many model parameters remain free even for this well-studied disk, the observed fluxes \emph{can} be reproduced to a level that is much better than a factor of two. We however refrain from trying to achieve so, as we feel that given the large uncertainties, these results would not be trustworthy and provide a misleading picture.

Given the considerable uncertainty and disagreement between the models, which of our conclusions is robust? From the considerably higher gas temperature needed to reproduce the high-$J$ CO lines, together with all PDR models agreeing that the gas temperature indeed decouples from the dust temperature at these high densities and UV fluxes, we conclude that there must be a warm gas atmosphere. We also consider the conclusions made in Section \ref{sec:paramstud} as robust, since they mainly represent \emph{relative} changes and trends that are more reliable than absolute fluxes, as long as the correct main heating and cooling processes dominate the thermal balance. Similarly, finding trends in the dependences of the line fluxes on model parameters, which can then be observationally tested using a larger sample of disks, is more promising than fitting the absolute fluxes. 

\subsection{The shape of the CO ladder} \label{sec:coshape}

\begin{figure}[htb]
\includegraphics[width=1.0\hsize]{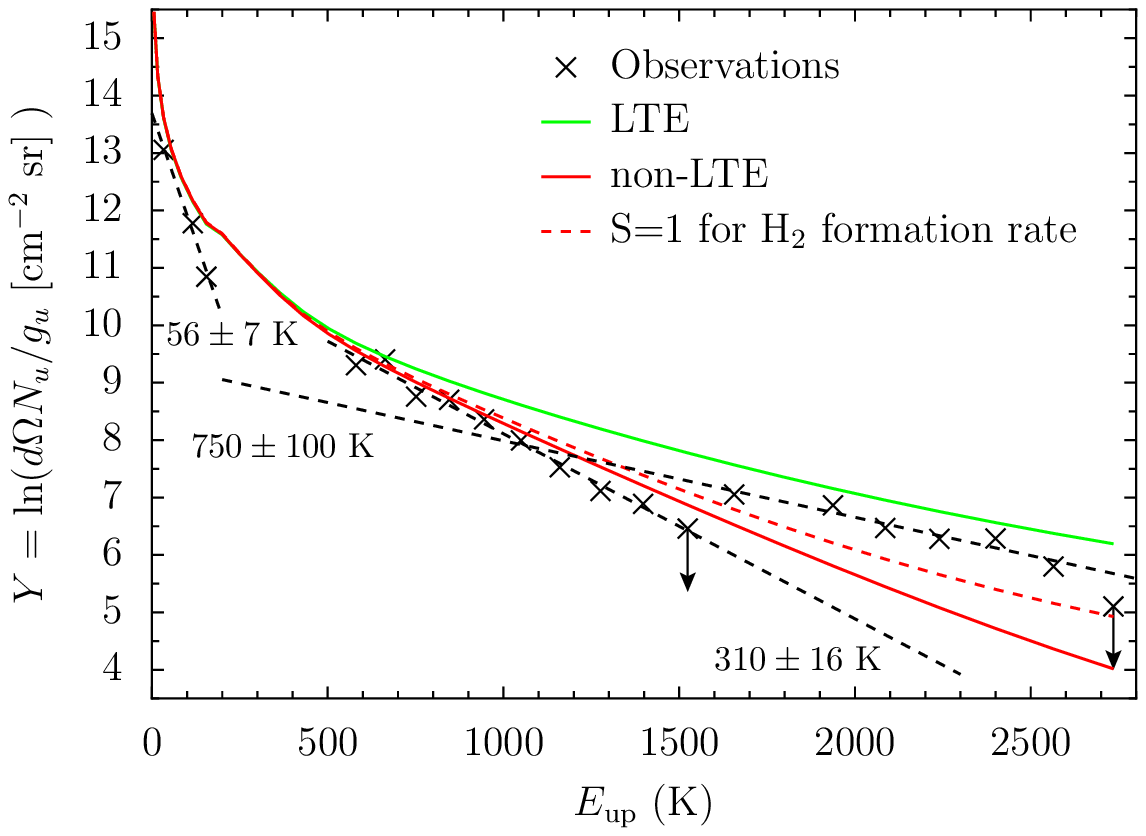}
\caption{Rotation diagram of the modeled and observed CO ladder.}
\label{fig:rotdiag}
\end{figure}

The high-$J$ CO lines emerge from the upper, inner atmosphere of the disk, from regions where heating by FUV radiation leads to a gas temperature that is much higher than the dust temperature (Section \ref{sec:lineorigin}). However, what does the shape of the CO ladder and in particular the mid-$J$ and high-$J$ CO lines tell us? In Figure \ref{fig:rotdiag}, we show the rotation diagram of the CO ladder from both observations and the representative model (Section \ref{sec:repmod}). \citet{Goldsmith99} provide a detailed introduction to the rotation diagram method. The $Y$ axis of the diagram is defined to include the solid angle of the emission, because the CO ladder is thought to emerge from different regions (Appendix \ref{sec:app_rotco}).

The observed CO ladder can be represented by three temperature components: A cold component with a rotation temperature of $T_{\rm rot} \sim 60$ K ($J_{\rm up} = 3 - 7$), a warm component of $T_{\rm rot} \sim 300$ K ($J_{\rm up} = 14 - 25$), and a hot component with $T_{\rm rot} \sim 750$ K ($J_{\rm up} = 25 - 30$). We did not include the upper limits to the $J=23-22$ and $J=31-30$ lines in the calculation of $T_{\rm rot}$.

The low-$J$ lines are optically thick. This leads to an increase in the flux compared to the optically thin conditions (Appendix \ref{sec:app_rotco}). Owing to the opacity effects, the temperature obtained from the cold component is thus not necessarily a kinetic temperature even though the lines are in thermal equilibrium. The model predicts that lines up to $E_u \sim 750$, $J=16-15$ are optically thick (Section \ref{fig:contriplot}). The modeled CO ladder indeed shows a curved shape, up to about this upper level energy, which is characteristic of optical depth effects. However, a larger emitting area, leading to a larger $d\Omega$ can also increase the line flux in a similar way. The emitting region $d\Omega$ contributing more than 50 \% to the total emission (obtained from the 75 \% radii given in Table \ref{tab:linorig}) increases from $J=16-15$ to $J=3-2$ by a factor of about eight, corresponding to a shift of $\Delta Y \sim 2$ in the rotation diagram.

The optically thin mid-$J$ and high-$J$ lines obtained from the model, can be represented well by a single temperature component of 393 K, which is somewhat higher than the observed value. The difference is however small in comparison to the large uncertainty in the gas temperature. The hot component of $T_{\rm rot} \sim 750$ K is not reproduced well by the representative model. A higher H$_2$ formation rate, leading to a higher H$_2$ abundance in the hot, inner gas, thus greater heating and more CO, yields an additional hot component. This model more close reproduces the observations of the hot component (Section \ref{sec:deph2}).

How is the CO ladder affected by the molecular excitation? To test this, we compared the CO ladder from the representative model with the same model, but assumed that the CO level population is in local thermal equilibrium (LTE). While the optically thick low-$J$ to mid-$J$ lines are very similar, the optically thin lines with higher $J$ are stronger by a factor of about nine ($\Delta Y \sim 2$) in flux. The critical density of these lines is on the order of up to a few times $10^6$ cm$^{-3}$ (Table \ref{tab:linorig}). In the region of the main emission (Section \ref{sec:lineorigin}), the density is similar or higher than the critical density. The calculated level population in this region is similar to the LTE population, to within $\sim 10$ \% for $J=16-15$ and to within a factor of two for $J=30-29$. The increase in the fluxes when assuming LTE is explained by the tenuous and hot gas at larger radii  also contributing to the emission, as in the case of for example the ``warm'' CO finger in the outer, upper disk, discussed in Section \ref{sec:chemstruct}.

We conclude that different effects govern the shape of the CO ladder including: \emph{(i)} The temperature of the emitting gas and the density. \emph{(ii)} Optical depth effects at low-$J$ to mid-$J$ lines. \emph{(iii)} For optically thin lines, the number of emitting molecules ($\sim N({\rm CO}) d\Omega$) and for optically thick lines, the emitting area ($\sim d\Omega$). Depending on the transition, different effects dominate. In addition, different temperature, column density, and density structures might reproduce the same CO ladder.

%
%
\section{Conclusions and outlook} \label{sec:conclus}

We have used a new detailed physical-chemical model to calculate the atomic fine-structure emission ([\ion{C}{I}], [\ion{C}{II}], and [\ion{O}{I}]) and CO lines from CO $J=3-2$ to $J=30-29$ from a protoplanetary disk around a young Herbig Ae star. We have applied the model to the well-studied HD 100546 disk, but the results are more generally applicable. We have compared our results to recent observations of the object and studied the dependence of the line flux on different parameters. Our main results are:
\begin{itemize}
\item The high-$J$ emission of CO can only be reproduced when the gas temperature in the inner disk is considerably higher than the dust temperature. A model with the gas temperature set to the dust temperature underpredicts the high-$J$ lines by orders of magnitude (Section \ref{sec:warmatmos}). The UV radiation is able to heat the gas sufficiently (Section \ref{sec:physstruc}).
\item Both models with a gas/dust $=100$ and $\delta_{\rm C}=0.05$ (volatile fraction of carbon $\delta_{\rm C}$) and models with gas/dust $=20$ and $\delta_{\rm C}=0.5$ are able to reproduce the CO ladder and the [\ion{O}{I}] lines. Models with a high gas/dust ratio overpredict the upper limit of [\ion{C}{I}] but approximately fit [\ion{C}{II}]. Models with a low gas/dust ratio, however agree with the upper limit of [\ion{C}{I}] but underpredict [\ion{C}{II}] (Section \ref{sec:paramgdrout}). 
\item None of the models reproduce [\ion{C}{I}]/[\ion{C}{II}], but models with higher gas/dust ratio have a lower ratio of these lines, closer to the observed value. The line ratio is also smaller in models with larger grains that allow UV radiation to penetrate deeper into the disk (Section \ref{sec:paramdust}, \ref{sec:cIcIIratio}) and models without PAHs in the outer disk, at radii $> 100$ AU. The ratio depends less on the stellar radiation field (Section \ref{sec:stespec}).
\item A substantial fraction of the [\ion{C}{II}] emission may emerge from a remnant envelope or a compact foreground (Section \ref{sec:cIIemission}), hence we prefer a solution with gas/dust ratio of 100, but a small fraction of volatile carbon (Section \ref{sec:carbbudget}). This means that a considerable part of the carbon is locked into more refractory carbon-bearing species and that the carbon partitioning between gas and dust has evolved from diffuse clouds to protoplanetary disk.
\item The highest-$J$ emission detected towards HD 100546, CO $J=30-29$, emerges from radii of $\sim 20-50$ AU. Mid-$J$ lines, e.g. CO $J=16-15$, are emitted at radii of $\sim 40-90$ AU and low-$J$ lines trace the outer disk (Section \ref{sec:lineorigin}).
\item The high-$J$ lines of CO ($J_{\rm up}>14$) are predicted to be much broader than the low-$J$ lines, which are observable from the ground (Section \ref{sec:colineshape}).
\item Changing the dust opacity law to very high values of $R_{\rm V}$ allows UV radiation to penetrate considerably deeper into the disk. The resulting line fluxes for a fixed gas/dust ratio are much larger. For 0.1-1.5 $\mu$m grains, gas/dust $=10$ with $\delta_{\rm C}=0.1$ reproduces the observations similarly well as gas/dust $=100$ and $\delta_{\rm C}=0.05$, if a $R_{\rm V}=5.5$ dust opacity law is used (Section \ref{sec:paramdust}).
\item Assuming a temperature-independent sticking coefficient for the H$_2$ formation only affects the high-$J$ lines, changing the line fluxes by a factor of a few. The uncertain H$_2$ formation rate is thus not a critical parameter for the CO emission, except for the highest-$J$ lines detected towards HD 100546 (Section \ref{sec:deph2}, \ref{sec:coshape}).
\item The shape of the CO ladder in the rotation diagram is governed by the combination of excitation effects, optical depth, and different emitting regions of the CO lines (Section \ref{sec:coshape}).
\end{itemize}

With this paper, we have introduced a new physical-chemical model and applied it to a protoplanetary disk. The model is based on our previous work (\citealt{Bruderer09b,Bruderer10}) and calculates the gas-temperature structure in a self-consistent way with the chemistry and the excitation of the molecules. The model has a wide range of applications and we plan to use it in the future to investigate both T Tauri disks as well as the outflow walls of the envelopes of YSOs, and to compare the results with Herschel observations. The model also can be relatively easily adopted to calculate three-dimensional structures and is well-suited to the analysis of upcoming high-resolution ALMA data.

\begin{acknowledgements}
We are grateful to Neal Evans, Ted Bergin, Gijs Mulders, Olivier Bern{\'e}, and the DIGIT team for stimulating discussions and scientific support. We thank the anonymous referee and Malcolm Walmsley for carefully reading the manuscript and useful suggestions. We thank Olja Pani{\'c} for providing the APEX data in electronic form. Astrochemistry in Leiden is supported by the Netherlands Research School for Astronomy (NOVA), by a Spinoza grant and grant 614.001.008 from the Netherlands Organisation for Scientific Research (NWO), and by the European Community's Seventh Framework Programme FP7/2007-2013 under grant agreement 238258 (LASSIE). The work on star formation at ETH Zurich is partially funded by the Swiss National Science Foundation grant 200020-113556. PACS has been developed by a consortium of institutes led by MPE (Germany) and including UVIE (Austria); KU Leuven, CSL, IMEC (Belgium); CEA, LAM (France); MPIA (Germany); INAF-IFSI/OAA/OAP/OAT, LENS, SISSA (Italy); IAC (Spain). This development has been supported by the funding agencies BMVIT (Austria), ESA-PRODEX (Belgium), CEA/CNES (France), DLR (Germany), ASI/INAF (Italy), and CICYT/MCYT (Spain).
\end{acknowledgements}

\bibliographystyle{aa}

\Online 

\begin{appendix}

\section{Details of the models and benchmark tests} \label{sec:app_detail}

\begin{figure}[htb]
\includegraphics[width=1.0\hsize]{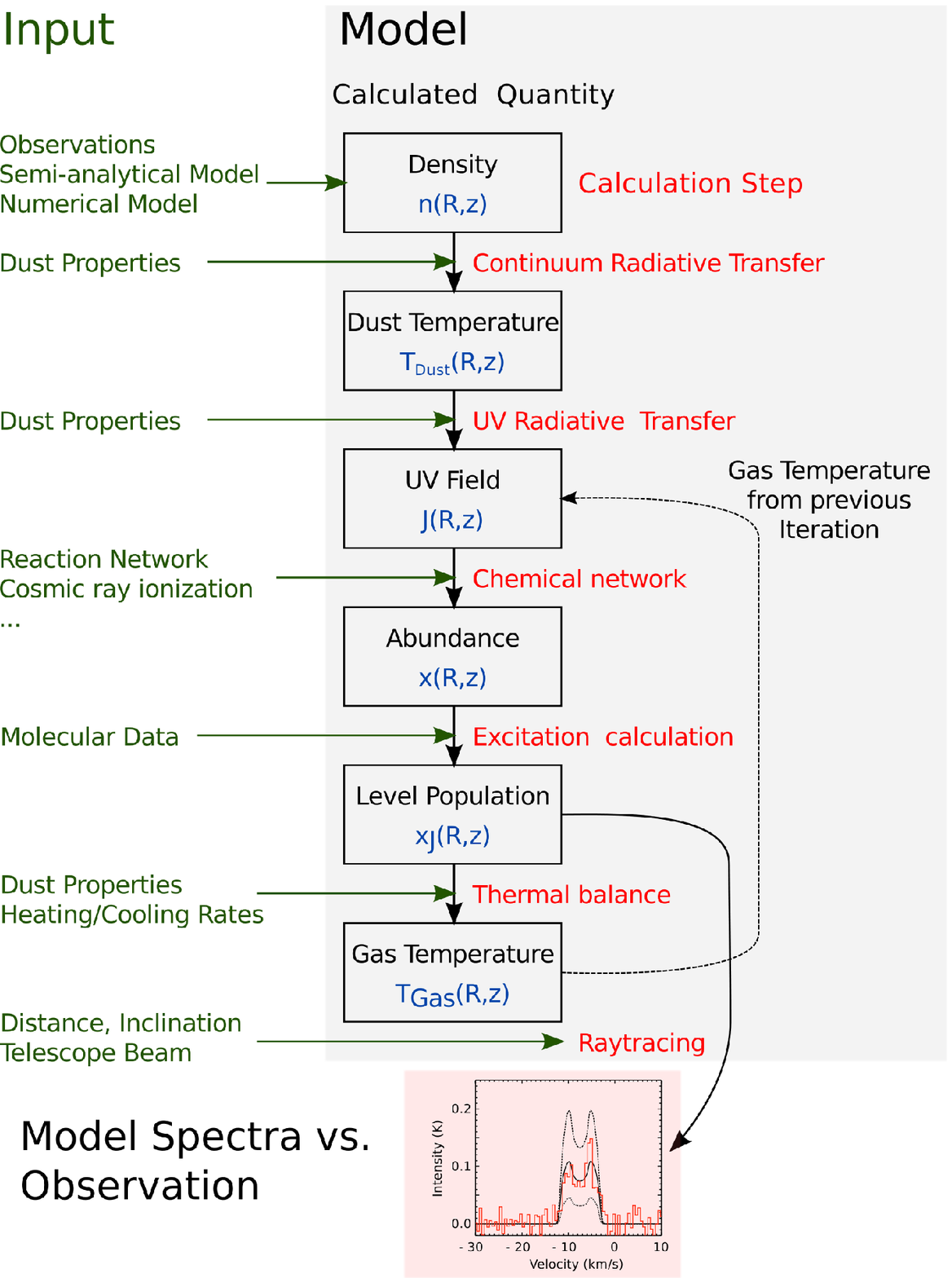}
\caption{Modeling flowchart}
\label{fig:modflow}
\end{figure}

The purpose of this appendix is twofold. First, it presents details of the implementation of the model and second, it compares of results with other codes and benchmark problems. The model is based on those of \cite{Bruderer09a,Bruderer09b,Bruderer10} and \cite{Bruderer10b}, although it contains several improvements. The structure of the model is summarized in Fig. \ref{fig:modflow}. The modeling process starts with a gas and density distribution e.g. obtained from an analytical prescription. Different modules then calculate the dust temperature and UV radiation (Sect. \ref{sec:moddustuv}), abundances of atomic or molecular species (Sect. \ref{sec:modchem}), and the level population of molecules or atoms acting as coolants (Sect. \ref{sec:modline}). The gas-temperature is then calculated from the balance between heating and cooling processes (Sect. \ref{sec:modthermal}). As the abundance and level population enter the heating and cooling rates, abundances and level populations need to be re-calculated until convergence. Finally, spectral lines can be calculated and compared with observations.

To carry out the calculation, the modeled region is divided into individual cells. In this present study, we assume a axisymmetric structure. Since all geometric information is contained in one separate module, the code can in principle also be run for spherically symmetric or three-dimensional structures.

The current study applies the model to a protoplanetary disk, but the model can also be used in the context of protostellar envelopes, for example to calculate the molecular emission from UV irradiated outflow walls. For future  applications, we thus present all modules in this appendix, even though they are not used in the current work (such as the calculation of the dust temperature).

\subsection{Dust radiative transfer} \label{sec:moddust}

\subsubsection{Dust temperature}

The dust temperature in steady state is obtained from the equilibrium between energy absorbed and emitted by dust
\begin{equation} \label{eq:dusteq}
\Gamma_{\rm abs} = \iint \kappa_\lambda I_\lambda d\lambda d\Omega = 4 \pi \int \kappa_\lambda B_\lambda(T_{\rm dust}) d\lambda =\Gamma_{\rm emit} \ ,
\end{equation}
where the dust opacity per unit mass is $\kappa_\lambda$, the incident radiation field is $I_\lambda$, and the Planck function for the dust temperature is $B_\lambda(T_{\rm dust})$. The radiation field consists of contributions from both the stellar radiation field and the ambient radiation by dust emission. Thus, different positions in the model are coupled by means of the radiative transfer equation. We neglect the ``back-warming'' of dust by hot gas and further assume a unique dust temperature for different sizes of dust.

In this present study, we implement the Monte Carlo technique proposed by \cite{Bjorkman01}: photon packages with an energy distribution following the stellar radiation field are propagated until interaction (scattering or absorption followed by re-emission). Scattering only changes the direction of a photon package. To sample the scattering angle, we use a Henyey-Greenstein function, which defines the scattering angle by only one parameter, $g=\langle \cos(\theta) \rangle$. Absorption increases the amount of energy deposited into a cell. For a cell of dust mass $M$, the dust temperature $T_{\rm new}$ after absorption of $N_i$ photon packages is given by
\begin{equation}
\int \kappa_\lambda B_\lambda(T^{\rm new}_{\rm Dust}) = \frac{N_i \delta L}{4 \pi M} ,
\end{equation}
with the energy per photon package $\delta L = L / N$, where the stellar luminosity $L$ is divided by the total number of photon packages $N$. When an absorption event increases the local temperature by $\Delta T=T^{\rm new}_{\rm Dust}-T_{\rm Dust}$, the photon-package is re-emitted at a wavelength given by (\citealt{Baes05})
\begin{equation}
P_\nu d\nu = \left(N_i+1\right) \frac{\kappa_\nu B_\nu(T_{\rm Dust}+ \Delta T) d\nu}{\int \kappa_\nu B_\nu(T_{\rm Dust}) d\nu} -
N_i \frac{\kappa_\nu B_\nu(T_{\rm Dust} + \Delta T) d\nu}{\int \kappa_\nu B_\nu(T_{\rm Dust}) d\nu} \ .
\end{equation}
This probability distribution function (PDF) accurately corrects for the incorrect spectrum of previously reemitted photons (at a too low temperature). However, it can sometimes become negative. In such cases, the approximative PDF given in \cite{Bjorkman01} is used. This procedure is generally very efficient as it does not require any global iteration and the dust temperature increases to the correct value. In regions of very high optical depth, however, photons can get stuck owing to a very high number of absorption and re-emission processes. We thus implement the diffusion approximation derived in \cite{Robitaille10}, based on \cite{Min09}. Regions of very low optical depth can suffer too little absorption resulting in a noisy dust temperature (see e.g. \citealt{Bruderer10b}, Fig. D.2). We implement a method proposed by \citet{Pinte06} and measure the ambient radiation field using the estimator given in \cite{Lucy99} during the propagation of the photons. After all photons have propagated, the dust temperature can be improved using this value. 

To benchmark the dust temperature calculation, we use the benchmark problems proposed in \cite{Ivezic97b} and compare our results with calculations obtained with the DUSTY code (\citealt{Ivezic97}). The benchmark problem consists of a spherical cloud with two different density gradients (either $\propto r^{-2}$ or constant) and four different total radial optical depth at 1 $\mu$m ($\tau_{1\ \mu m}=1, 10, 100$, or $1000$). The input spectra is a black body and the dust properties follow a simple analytical law. Scattering is assumed to be isotropic. The agreement between the codes is perfect, being closer than about a percent.

\begin{figure}[htb]
\includegraphics[width=0.8\hsize]{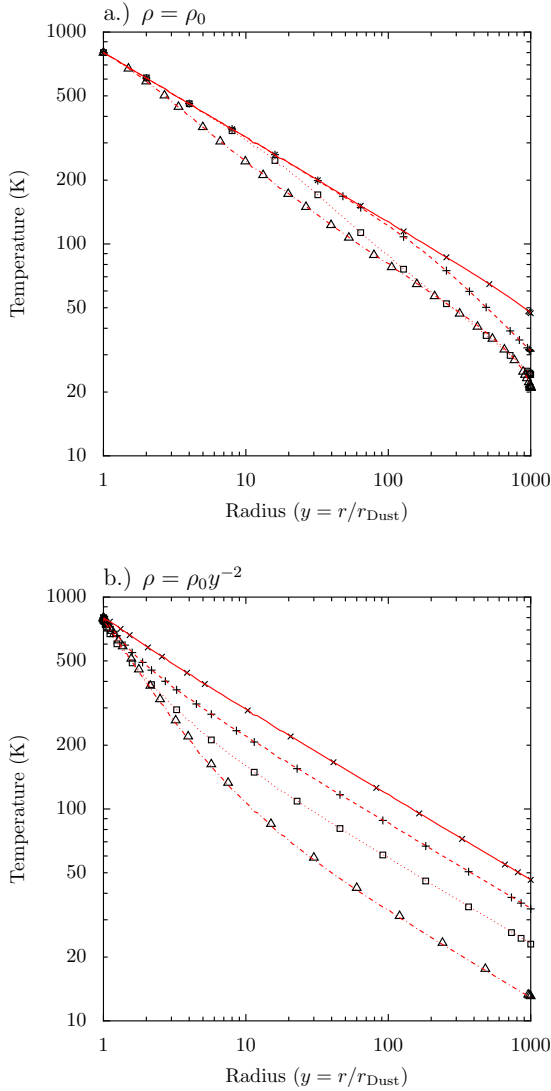}
\caption{Results of a spherical benchmark test for the dust temperature calculation (\citealt{Ivezic97b}). The dust temperature obtained from this work is given by a red line, while the results from DUSTY are given by black signs. The solid line corresponds to $\tau_{1 \mu m}$, and the dashed, dotted, and dash-dotted lines to $\tau_1=10, 100$, and $1000$, respectively. a.) constant density, b.) density $\propto r^{-2}$.
\label{fig:benchmark_tdust_dusty}}
\end{figure}

\begin{figure}[htb]
\includegraphics[width=\hsize]{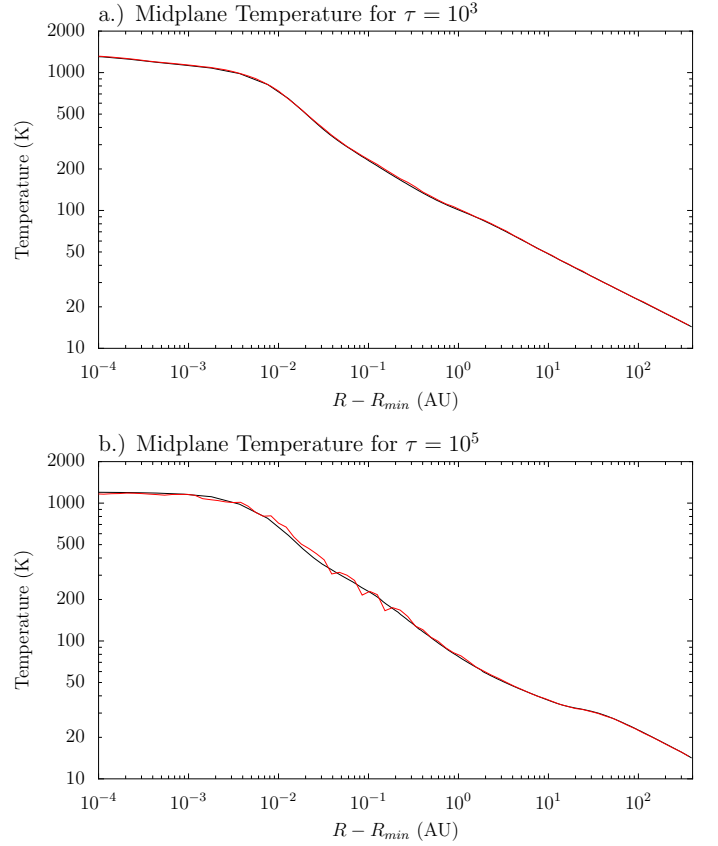}
\caption{Dust temperature at midplane for the disk benchmark problem (\citealt{Pinte09}). The result from our code is given in red, the result from MCFOST (\citealt{Pinte06}) in black.}
\label{fig:benchmark_tdust_disk}
\end{figure}

As a second benchmark problem, we calculate the dust temperature in a protoplanetary disk with very high total optical depth (\citealt{Pinte09}). We calculate their benchmark problem for isotropic scattering with a midplane optical depth at 0.86 $\mu$m of $\tau=10^3$ and $10^5$ (their Fig. 9). In Fig. \ref{fig:benchmark_tdust_disk}, we show the midplane temperature for the two models compared to the results obtained with MCFOST. The agreement is very good for the model with $\tau=10^3$. For the run with $\tau=10^5$, the agreement is again very good with deviations smaller than a few percent,  even in the region shadowed by the dense inner rim. These deviations are comparable to those between the other codes participating in that benchmark study.

\subsection{UV field} \label{sec:moddustuv}

Since the calculation of the dust temperature does not explicitly compute the radiation field as a function of  wavelength, the intensity at UV and optical wavelengths is calculated in a second step. This intensity is used for example to calculate photodissociation rates. The calculation of the intensity uses the same method for the propagation of the photon packages as the calculation of the dust temperature. The radiation field is then calculated with the method proposed by \cite{vanZadelhoff03}. For each wavelength bin, $N$ photon packages are propagated. All photon packages are assigned with an initial intensity $I_\lambda(0) = F_\lambda S/ N$, where $F_\lambda$ is the input flux and $S$ the surface it passes through. This intensity then drops to $I_\lambda(s+\Delta s) = I_\lambda(s) e^{-\Delta \tau_\lambda}$, when the package advances by $\Delta s$ with an optical depth $\Delta \tau_\lambda$. The mean intensity is calculated from 
\begin{equation}
J_\lambda = \frac{1}{4\pi V} \sum_i I_{\lambda,i} \Delta s_i \frac{1- e^{-\Delta \tau_{\lambda,i}}}{\Delta \tau_{\lambda,i}} \ ,
\end{equation}
with the volume of the cell $V$ and the sum taken over all photon packages passing through a cell. The distance and optical depth as an individual photon package $i$ passes through a cell are given by $\Delta s_i$ and $\Delta \tau_{\lambda,i}$. Since we calculate the mean intensity for a few wavelengths only, the input intensity is obtained by averaging over the input (stellar) spectra in each individual spectral band. Photons from the interstellar radiation field (\citealt{Draine78}) are also accounted for.

As a benchmark test, we calculate the mean intensity in a spherical cloud, switching scattering off. We compare the calculated intensity to the analytically obtained intensity and find agreement to better than one percent. The implementation of the scattering follows the calculation of the dust temperature and was tested in the previous paragraph.

\subsection{Chemistry} \label{sec:modchem}

Abundances of molecular and atomic species are obtained from the solution of the rate equations
\begin{equation} \label{eq:chemrates}
\frac{d n(i,t)}{dt} = \sum_j k_{ij} n(j,t) + \sum_{jl} k_{ijl} n(j,t) n(l,t) \ ,
\end{equation}
with the abundance $n(i,t)$ (cm$^{-3}$) of a species $i$ at time $t$. The rate of destruction and formation of a species is given by the coefficients $k_{ij}$ and $k_{ijl}$, respectively. In addition, three-body reactions can be entered, but our current network does not contain any. The set of coupled, stiff, and non-linear differential equations \ref{eq:chemrates} is solved by the DVODE solver (\citealt{Brown89}) for the temporal evolution of the species. We verify that the solution complies with the conservation of elements and charge. We note that the DVODE solver is widely used for applications in astrochemistry and well-tested (e.g. \citealt{Nejad05}). To obtain steady-state abundances, Eq. \ref{eq:chemrates} is solved for $d n(i,t) / dt = 0$ using a globally convergent Newton-Raphson method. To close the system of equations Eq. \ref{eq:chemrates}, some rate equations are replaced by conservation equations. If the steady-state solver fails to converge, the abundances are obtained from the time-dependent solver running to a high chemical age ($>10$ Mio. years). 

\begin{figure*}[htb]
\includegraphics[width=\hsize]{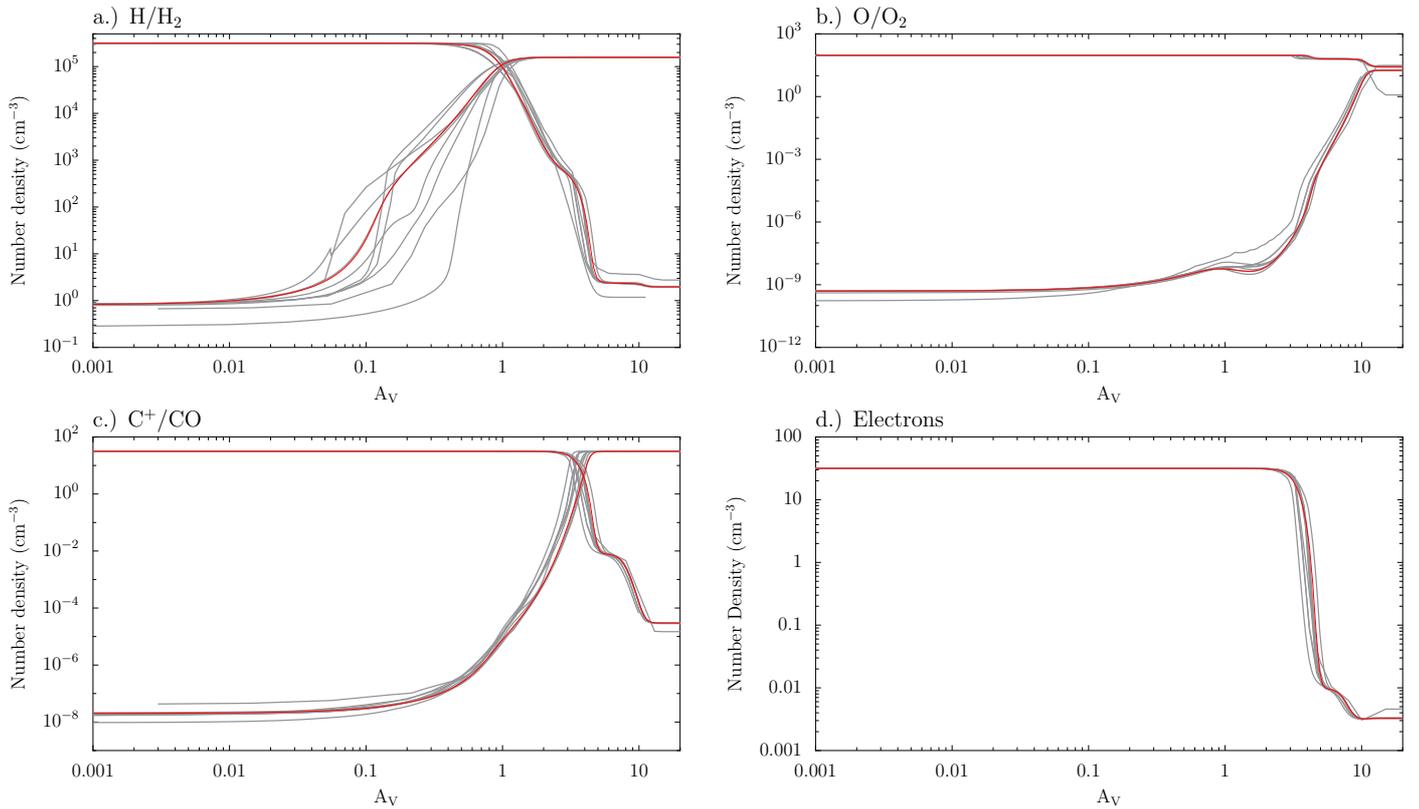}
\caption{Result of the chemistry benchmark for a slab with UV irradiation (\citealt{Roellig07}, problem F4). The result of our code is given by the red line, the results of the PDR codes by gray lines.}
\label{fig:benchmark_chempdr}
\end{figure*}

To test numerical aspects of the solver, we calculate abundances in an isothermal slab of constant density with UV irradiation. A small chemical network from the PDR comparison study by \cite{Roellig07} is used for this test. We first calculate their problem F4 (for a density of $10^{5.5}$ cm$^{-3}$, UV field of $\chi=10^5$, and a temperature of 50 K) with the steady-state solver. In Fig. \ref{fig:benchmark_chempdr}, we give the abundances of H, H$_2$, O, O$_2$, C$^+$, CO, and electrons and compare them to the results obtained from PDR codes. Our code to within their scatters agrees with the other codes. The scatter in the H$_2$ abundance is due to a different implementation of some reactions, such as H$_2$ dissociation including self-shielding. We also calculated the abundances with the time-dependent solver running to an old chemical age and found an agreement to better than a factor of $10^{-4}$ with the steady-state solution.

\subsubsection{Chemical network} \label{sec:chemnet}

The chemical network used in our work contains 10 elements, 109 species, and 1492 reactions. We use the same elemental composition as \cite{Jonkheid06} and \cite{Woitke09}, which is summarized in Table \ref{tab:chemelem}. The species contained in the network are given in Table \ref{tab:chemspec}. We note that H$_2^*$ is vibrationally excited H$_2$. PAH$^0$, PAH$^+$, and PAH$^-$ are neutral, positively, and negatively charged PAHs respectively and PAH:H denotes hydrogenated PAHs. Species frozen-out onto dust grains are shown by JX, for example JCO is CO frozen-out.

\begin{table}[tbh]
\caption{Elemental composition. $a(b)$ means $a \times 10^b$.}
\label{tab:chemelem}
\centering
\begin{tabular}{ll}
\hline\hline
Element & number fraction \\
\hline
H         &      1        \\
He        &      7.59(-2) \\
C         &      2.70(-4)$\delta_{\rm C}$\tablefootmark{a} \\
N         &      2.14(-5) \\
O         &      2.88(-4) \\
Mg        &      4.17(-6) \\
Si        &      7.94(-6) \\
S         &      1.91(-6) \\
Fe        &      4.27(-6) \\
\hline
\end{tabular}
\tablefoot{\tablefoottext{a}{depending on the amount of volatile carbon, see main text.}}
\end{table}

\begin{table}[tbh]
\caption{Species contained in the chemical network. See Section \ref{sec:chemnet} for an explanation of the species.}
\label{tab:chemspec}
\centering
\begin{tabular}{cccccc}
\hline\hline
H         &He        &C         &N            &O         &Mg           \\
Si        &S         &Fe        &H$_2$        &H$_2^*$   &CH           \\
CH$_2$    &NH        &CH$_3$    &NH$_2$       &CH$_4$    &OH           \\
NH$_3$    &H$_2$O    &CO        &HCN          &HCO       &NO           \\
H$_2$CO   &O$_2$     &HS        &H$_2$S       &CO$_2$    &SO           \\
OCS       &CN        &N$_2$     &SiH          &CS        &HCS          \\
SO$_2$    &SiO       &H$_2$CS   &H$^+$        &H$^-$     &H$_2^+$      \\
H$_3^+$   &He$^+$    &HCO$^+$   &C$^+$        &CH$^+$    &N$^+$        \\
CH$_2^+$  &NH$^+$    &CH$_3^+$  &O$^+$        &NH$_2^+$  &CH$_4^+$     \\
OH$^+$    &NH$_3^+$  &CH$_5^+$  &H$_2$O$^+$   &H$_3$O$^+$&Mg$^+$       \\
CN$^+$    &HCN$^+$   &Si$^+$    &CO$^+$       &HCNH$^+$  &SiH$^+$      \\
NO$^+$    &SiH$_2^+$ &S$^+$     &O$_2^+$      &HS$^+$    &H$_2$S$^+$   \\
H$_3$S$^+$&SiO$^+$   &CS$^+$    &CO$_2^+$     &HCS$^+$   &SO$^+$       \\
Fe$^+$    &SO$_2^+$  &HSO$_2^+$ &SiOH$^+$     &H$_2$CS$^+$&H$_3$CS$^+$ \\
HSO$^+$   &OCS$^+$   &HOCS$^+$  &S$_2^+$      &HN$_2^+$  &HS$_2^+$     \\
e$^-$     &PAH$^0$   &PAH$^+$   &PAH$^-$      &PAH:H     &JC           \\
JN        &JO        &JCH       &JCH$_2$      &JNH       &JCH$_3$      \\
JNH$_2$   &JCH$_4$   &JOH       &JNH$_3$      &JH$_2$O   &JCO          \\
JCO$_2$   &          &          &             &          &             \\
\hline
\end{tabular}
\end{table}

The chemical reaction network is based on the studies by \cite{Doty04}, \cite{Staeuber05}, and \cite{Bruderer09a}, but contains some extensions. The following reaction types are included:

\subsubsection*{1.) H$_2$ formation on dust}

For the formation of H$_2$ on dust grains, we implement the rate by \cite{Sternberg89}, scaled to the appropriate grain size. We use the sticking coefficients of \citet{Cuppen10}. Since the gas/dust ratio can vary over a model, rates involving small dust are scaled to the local gas/dust ratio.

\subsubsection*{2.) Freeze-out, thermal, and non-thermal desorption}

Freeze-out and thermal or non-thermal desorption are implemented following \cite{Visser09c} and references therein. The thermal desorption rates account for the abundance of the ice on the surface and change from zeroth to first order behavior when the ice contains less than a monolayer of ice. As in \cite{Visser09c}, we use the same pre-exponential factor for all species except for H$_2$O (\citealt{Fraser01}) and CO (\citealt{Bisschop06}). Binding energies follow \cite{Aikawa97} and \citet{Sandford93}. 

Non-thermal evaporation by UV photons or cosmic-rays depends on a yield indicating the number of molecules desorbed per grain per incident UV photon. We assume the same yields used by \citet{Visser09c}, following \citet{Oberg07,Oberg09a}. To simulate cosmic-ray induced desorption, a small background UV flux corresponding to $10^4$ photons cm$^{-2}$ s$^{-1}$ (\citealt{Shen04}) is added to the stellar UV flux.

\subsubsection*{3.) Hydrogenation of simple species on ices}

In addition to freeze-out and evaporation, we implement the grain-surface hydrogenation of simple species. We follow the approach of \cite{Visser09c} and \cite{Hollenbach09}. The hydrogenation leads to saturation of species frozen-out, from JC $\rightarrow$ JCH $\rightarrow$ JCH$_2$ $\rightarrow$ JCH$_3$ $\rightarrow$ JCH$_4$, JN $\rightarrow$ JNH $\rightarrow$ JNH$_2$ $\rightarrow$ JNH$_3$and JO $\rightarrow$ JOH $\rightarrow$ JH$_2$O. Other grain-surface reactions to build up more complex species are not accounted for, but the focus of our work is on simple species.

\subsubsection*{5.) Gas-phase reactions}

Most reactions in our network (1132 out of 1492) are pure gas-phase reactions, such as ion-neutral, neutral-neutral, or charge exchange reactions. Our network contains all reactions from the UMIST 2006 database (\citealt{Woodall07}) involving the species given in Table \ref{tab:chemspec}. Details of the implementation are given in \cite{Bruderer09a}.

\subsubsection*{6.) Photodissociation}

Photodissociation rates are calculated from the local radiation field using the cross-sections for discrete and continuous absorption given by \cite{vanDishoeck88}, \cite{vanDishoeck06c}, and \cite{vanHemert08}\footnote{www.strw.leidenuniv.nl/\~{}ewine/photo/}. For the dissociation of H$_2$, CO, and the ionization of C, we include the effects of self-shielding by reducing the unshielded rate with a self-shielding factor. For H$_2$, the factor given in \cite{Draine96} is used. For CO, we implement the factors given in \cite{Visser09b} and for C, the factor of \cite{Kamp00} is used. We however limited the shielding of C by H$_2$ to 0.5, owing to the overlap of H$_2$ lines with the wavelengths at which carbon is photoionized.

The self-shielding rates depend on the column densities of H$_2$, C, and CO towards the UV source. We calculate these column densities together with the UV field (Sect. \ref{sec:moddustuv}) by averaging over the densities of all photon packages passing through a cell, weighted by their intensity.

\subsubsection*{7.) X-ray induced processes}

As demonstrated by \cite{Bruderer09a}, the influence of X-rays on the chemistry can be accurately approximated by secondary ionizations induced by fast photo-electrons from H$_2$ and H ionization. We follow their approach and calculate the X-ray ionization rate using the cross-sections given in their Appendix A. The rates for the secondary processes are taken from \cite{Staeuber05}. We note that the cosmic-ray-induced photodissociation rates are scaled-up to the X-ray ionization rate following \cite{Staeuber05}.

\subsubsection*{8.) Cosmic-ray-induced reactions}

Rates for direct cosmic-ray-induced ionization processes are taken from the UMIST 2006 network. Cosmic-ray-induced photoionization rates are also taken from that database, with the exception of CO, H$_2$, and H, where we follow \cite{Staeuber05}. For the CO dissociation, we use the rates of \cite{Gredel87} implemented with the fitting equation provided in \cite{Maloney96}. We account for the ionization of CO, H$_2$, and H by the decay of excited He (2${}^1$P state, 19.8 eV, see \citealt{Yan97}).

\subsubsection*{9.) PAH/small grain charge exchange/hydrogenation}

We assume that PAHs and small grains are well-mixed with the gas and thus not scaled to the gas/dust ratio. For the photoionization, charge-exchange, and recombination, we use the rates given in \citet{Wolfire03} with a collision rate parameter $\phi_{\rm pah} = 0.5$. For species heavier than hydrogen, the recombination rates are scaled to $1/\sqrt{m_{\rm amu}}$, where $m_{\rm amu}$ is the mass of the molecule. 

The formation of CH$^+$ and H$_2$ on PAHs is implemented using the rates of \cite{Jonkheid06}. We include the reactions 
\begin{eqnarray}
{\rm H} + {\rm PAH}^0 &\rightarrow& {\rm PAH:H} \label{eq:pahh0}\\
{\rm H}^+ + {\rm PAH}^- &\rightarrow& {\rm PAH:H} \label{eq:pahh1} \\
{\rm H} + {\rm PAH:H} &\rightarrow& {\rm PAH}^0 + {\rm H}_2 \label{eq:pahh2}\\
{\rm C}^+ + {\rm PAH:H} &\rightarrow& {\rm PAH}^0 + {\rm CH}^+ \label{eq:pahh3} 
\end{eqnarray}
Reactions \ref{eq:pahh0}, \ref{eq:pahh1}, and \ref{eq:pahh2} run at $10^{-9}$ cm$^{3}$ s$^{-1}$, while \ref{eq:pahh3} runs at the same speed as the recombination of C$^+$ with PAH$^0$. 

\subsubsection*{10.) Reactions with H$_2^*$ (excited state H$_2$)}

Vibrationally excited H$_2$ is included into the chemistry with a two-level approximation. H$_2^*$ denotes H$_2$ in a vibrationally excited pseudo-level with an energy corresponding to 30163 K (\citealt{London78}). The energy stored in the vibrational state can be used to overcome an activation barrier (e.g. \citealt{Sternberg95}). Following \cite{Tielens85}, the rate coefficients of a species with H$_2^*$ can be approximated using the rate for H$_2$ by reducing the exponential factor by 30163 K. For a H$_2$ rate $\propto \exp(-\gamma/T_{\rm kin})$, the exponential factor for H$_2^*$ is thus $\gamma^*=\max(0,\gamma-30163 {\rm K})$. This approximation however overestimates the rate coefficients, if $\gamma$ is too large and we switch off reactions with $\gamma > 20000$ K. 

The population of H$_2^*$ is determined by collisions with H and H$_2$ (\citealt{Tielens85}), spontaneous decay ($2\times10^{-7}$ s$^{-1}$; \citealt{London78}) and the pumping by FUV photons, which is assumed to be ten times the photodissociation rate.

\subsection{Molecular/atomic excitation} \label{sec:modline}

The population $n_i$ (in cm$^{-3}$) of a level $i$ of a molecular or atomic species is determined by the rate equation
\begin{equation} \label{eq:linert_rate}
\frac{d n_i}{dt} = \sum_{i \neq j} n_j P_{ij} - n_i \sum_{i \neq j} P_{ij} + F_i - D_i = 0 , 
\end{equation}
with the rate coefficients for the transition between level $i$ and $j$,
\begin{equation}
P_{ij} = \left\{
\begin{array}{ll}
A_{ij}+B_{ij} \langle J_{ij} \rangle + C_{ij} & (E_i > E_j) \\
B_{ij} \langle J_{ij} \rangle + C_{ij} & (E_i < E_j) \ , \\
\end{array}
\right.
\end{equation}
with the level energy $E_i$, the collisional rate coefficients $C_{ij}$ and the Einstein A and B-coefficients $A_{ij}$ and $B_{ij}$, respectively. The rate of chemical formation of a molecule forming into the state $i$ and destroyed from that state is given by $F_i$ and $D_i$, respectively (\citealt{Staeuber09}). In the following, we assume steady-state conditions and set $d n_i / dt = 0$. To calculate the ambient radiation field
\begin{equation} \label{eq:linert_jmean}
\langle J_{ij} \rangle = \frac{1}{4 \pi} \iint \phi_{ij}(\nu) I_\nu(\Omega) d\nu d\Omega \ ,
\end{equation}
the normalized line profile function $\phi_{ij}(\nu)$ and the intensity depending on the frequency and direction is needed in every cell of the model. The intensity along a ray and for one frequency is calculated from the radiative transfer equation 
\begin{equation} \label{eq:linert_radi}
\frac{dI_\nu}{ds} = - \alpha_\nu I_\nu + j_\nu \ ,
\end{equation}
with the absorption and emission coefficient $\alpha_\nu$ and $j_\nu$, respectively and the distance $ds$ that the ray propagates. Both coefficients contain contributions from both line and dust absorption and emission. Since the level population enters the absorption and emission coefficients, Eqs. \ref{eq:linert_rate}, \ref{eq:linert_jmean}, and \ref{eq:linert_radi} form a coupled problem, which is computationally very demanding to solve (see for example \citealt{Hogherheijde00,Brinch10}). 

In this work, we employ an approximation similar to the escape probability method (e.g. \citealt{vdTak07}). In this approximation, the physical conditions and thus the source function $S_\nu = j_\nu / \alpha_\nu$ is assumed to be constant in the modeled region. The ambient radiation field $\langle J_{ij} \rangle$ can then be expressed by an analytical expression, which greatly facilitates the calculation. For our models with physical conditions strongly varying with position, the approximation of a single zone with constant physical conditions is not applicable. The solution of Eq. \ref{eq:linert_radi} then reads
\begin{equation} \label{eq:linert_solveradi}
I_\nu = I_{\nu,{\rm bg}} e^{-\tau_\nu} + \int_0^{\tau_\nu} S_\nu(\tau_\nu') e^{-(\tau_\nu-\tau_\nu')} d\tau_\nu' \ ,
\end{equation}
using the definition of the optical depth, $d\tau_\nu = \alpha_\nu ds$ and the background intensity $I_{\nu,{\rm bg}}$. Similar to \cite{Poelman05}, we solve for the excitation at every position, although we approximate the local radiation field by keeping the source function constant along a ray. The time-consuming integration in Eq. \ref{eq:linert_solveradi} reduces in this way to a simple calculation of the opacity along a ray from the current position to the edge of the modeled region. This approximation allows us to solve for the excitation of complex problems (many lines, high optical depth) in a short time. For example the excitation of water in the models presented in this work can be solved in a few minutes using a standard PC. 

To include pumping by dust and the velocity structure, we follow the approach of \cite{Takahashi1983}. For the dust and line opacity $\tau_{\nu,{\rm L}}$ and $\tau_{\nu,{\rm D}}$, along a ray, we calculate
\begin{eqnarray}
\epsilon_{ij} &=& \frac{1}{4\pi} \iint \frac{\tau_{\nu,{\rm D}}+\tau_{\nu,{\rm L}} e^{-(\tau_{\nu,{\rm L}}+\tau_{\nu,{\rm D}})}}{\tau_{\nu,{\rm D}}+\tau_{\nu,{\rm L}}} \phi_{ij}(\nu) d\nu d\Omega \\
\eta_{ij} &=& \frac{1}{4\pi} \iint e^{-(\tau_{\nu,{\rm L}}+\tau_{\nu,{\rm D}})} \phi_{ij}(\nu) d\nu d\Omega \ .
\end{eqnarray}
The level population is then obtained from Eq. \ref{eq:linert_rate} with
\begin{equation}
P_{ij} = \left\{
\begin{array}{ll}
A_{ij}\epsilon_{ij}+B_{ij} \langle J'_{ij} \rangle + C_{ij} & (E_i > E_j) \\
B_{ij} \langle J'_{ij} \rangle + C_{ij} & (E_i < E_j) \ , \\
\end{array}
\right.
\end{equation}
and $\langle J'_{ij} \rangle = (\epsilon_{ij} -Ê\eta_{ij}) B_\nu(T_{\rm D}) + \eta_{ij} I_{\nu,{\rm bg}}$. The local dust temperature is given by $T_{\rm D}$ and the background radiation (CMB) is given by $I_{\nu,{\rm bg}}$. For the calculation of $\epsilon_{ij}$ and $\eta_{ij}$, either 6 or 26 directions and of order 40 frequency bins per line are used. 

Once the level population is determined, the molecular cooling rate can be calculated and synthetic maps can be derived by solving the radiative transfer equation (Eq. \ref{eq:linert_solveradi}). Our raytracer produces images in fits format. To properly resolve the inner structure (hot core or inner disk), we shoot several rays per pixel through that region and average over the intensity. Along the ray, the integration steps within one computational cell are refined, according to the velocity structure in order to properly resolve narrow lines.

\subsubsection*{Benchmark test}

\begin{figure}[htb]
\includegraphics[width=\hsize]{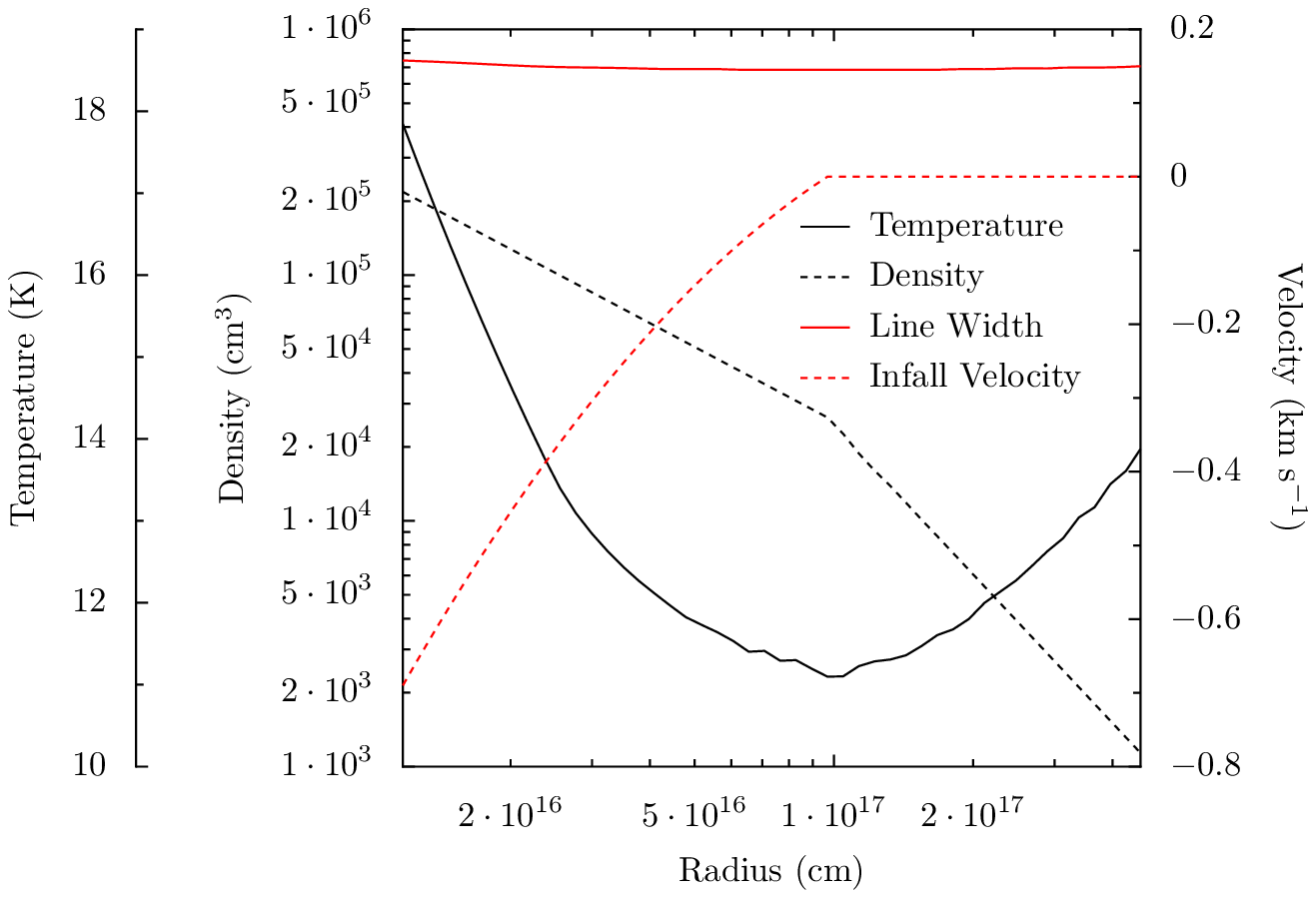}
\caption{Physical conditions of the cloud core used to benchmark the excitation calculation.}
\label{fig:benchmark_hcop_struct}
\end{figure}

To benchmark the excitation calculation, we calculate the problem proposed by \citet{vanZadelhoff02}, which consists of the calculation of the HCO$^+$ excitation in a spherical, infalling class 0 cloud core. The physical conditions are given in Fig. \ref{fig:benchmark_hcop_struct}. The abundance of HCO$^+$ is fixed to either $10^{-8}$ or $10^{-9}$ relative to the H$_2$ density. The difficulty of this benchmark problem lies in the high optical depth reached in the lines and that collision partner densities are below the critical density. The level population is thus far from the local thermal equilibrium and radiative excitation is important.

\begin{figure*}[tbh]
\includegraphics[width=1.0\hsize]{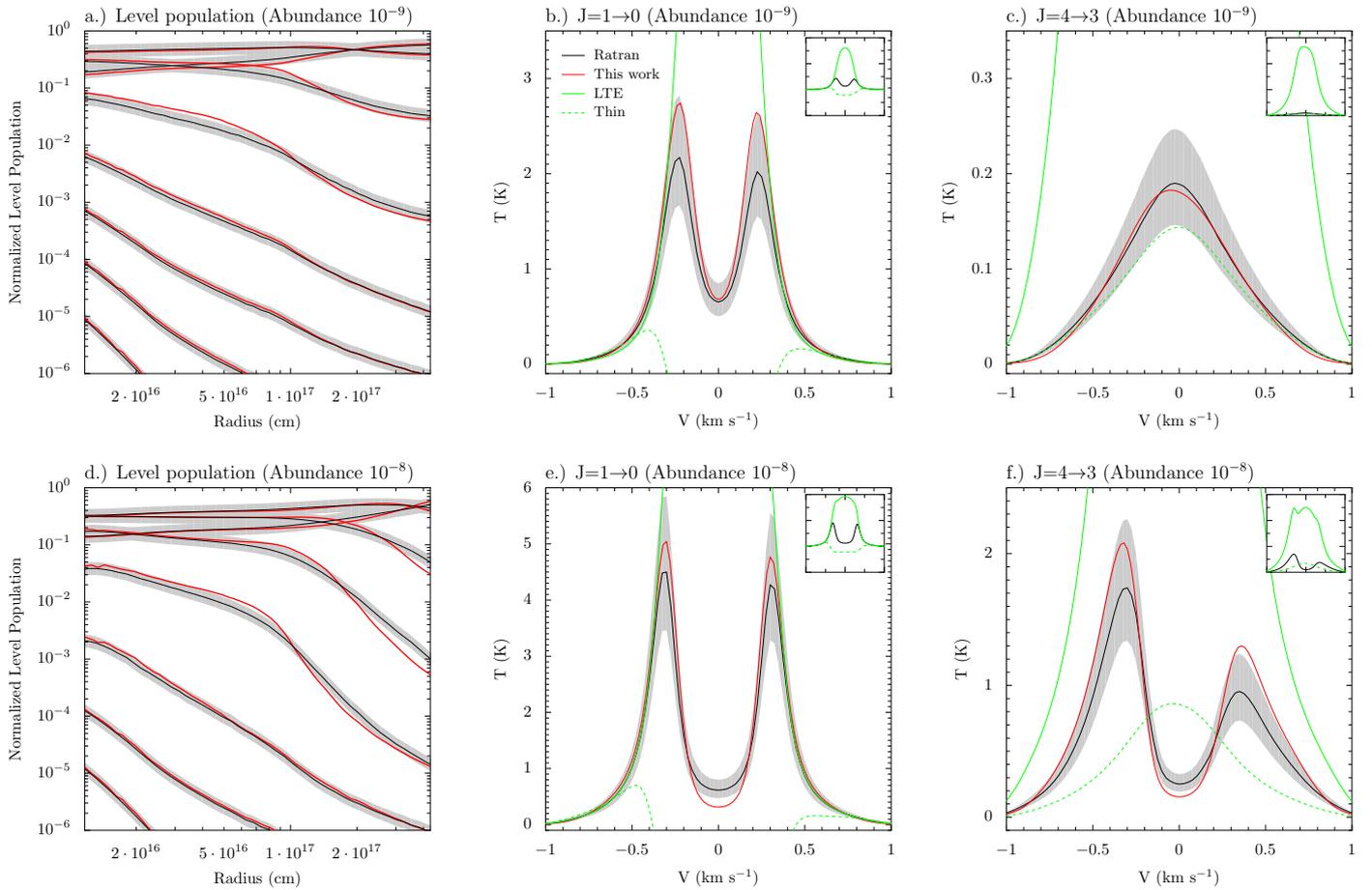}
\caption{Result of the benchmark of the excitation calculation. The upper panel shows the results for a HCO$^+$ abundance of $10^{-9}$ relative to H$_2$, the lower panel for an abundance of $10^{-8}$ relative to H$_2$. The plots on the left display the normalized level population of the first eight levels, the plots on the right show beam-averaged intensities. The black lines give results obtained using the RATRAN code, and the red dots/lines show results derived by this work. The gray shaded area indicates a 30\% range to the RATRAN solution.\label{fig:benchmark_hcop}}
\end{figure*}

In Fig. \ref{fig:benchmark_hcop}, we compare the solution obtained with this method with the solution calculated by RATRAN (\citealt{Hogherheijde00}). That code is widely used and has been well-tested against other codes. For an abundance of $10^{-9}$, the normalized level population agrees to within 30\% with the solution provided by RATRAN. An exception is the $J=3$ level in the inner part of the cloud. The deviations are larger for an abundance of $10^{-8}$, but still mostly within 30\%. The intensity is obtained with the raytracer implemented in our code and convolved to the telescope beam. In Fig. \ref{fig:benchmark_hcop}, the $J=1\rightarrow0$ and $J=4\rightarrow3$ lines are shown for a beam of $29''$ and $14''$, respectively. We also indicate the intensities obtained assuming local thermal equilibrium (LTE) and ``thin'' conditions, setting the ambient radiation field to 0. As the level populations, the line fluxes mostly agree to within about 30\% of the intensity obtained with RATRAN. We note that the intensities obtained with RATRAN are calculated with their raytracer (SKY) and convolved using the MIRIAD package (\citealt{Qi05}). Thus, this benchmark problem also tests our raytracer and convolution routine.

The agreement found in with this benchmark problem is similar to the findings of \cite{Woitke09a}, who calculated the water emission from a Herbig Ae disk calculated with a similar method and compared them to RATRAN. We consider the agreement as reasonable and sufficient for our application, since the uncertainties entering through the chemical network calculation are much larger. It also shows that this method is a much better approximation than e.g. assuming LTE. 

\subsection{Thermal balance} \label{sec:modthermal}

The gas temperature is obtained from the equilibrium between the heating and cooling rates
\begin{equation}
\frac{d \epsilon}{dt} = \sum_i \Gamma_i - \sum_i \Lambda_i = 0 , 
\end{equation}
where $\epsilon$ is the internal energy of the gas and $\Gamma_i$ and $\Lambda_i$ are different heating and cooling rates (for example, line cooling or photoelectric heating on dust grains). This non-linear equation is solved by a secant method starting from the dust temperature. If the convergence is insufficient, the process is restarted with a different value. Convergence is reached if the heating and cooling rate agree to within a predefined threshold in all cells (typically 1\%). We implement the following heating and cooling rates (see \citealt{Bruderer09b}):

\subsubsection*{1.) Photoelectric heating}

Photoelectric heating on large graphite and silicate grains is implemented following \cite{Kamp01}. The required grain absorption cross-section, averaged over the FUV band, is taken to be consistent with the calculation of the UV field and the dust temperature. As for the chemistry, the rates involving dust grains are scaled to the local gas/dust ratio. Photoelectric heating on small dust grains is implemented following \cite{Bakes94}. We also implement their recombination cooling rate.

\subsubsection*{2.) Gas-grain heating or cooling}

Depending on the difference between gas and dust temperature, gas-grain collisions can either heat or cool the gas. In regions of very high density, gas-grain collisions couple the gas temperature to the dust temperature (e.g. \citealt{Doty97a}). We implement the heating and cooling rates following \cite{Young04}.

\subsubsection*{3.) H$_2$ heating}

Molecular hydrogen can contribute to the heating and cooling in different ways: \textit{(i)} Line cooling. \textit{(ii)} Heating through collisional deexcitation of FUV pumped H$_2$. \textit{(iii)} Formation heating. \textit{(iv)} Photodissociation heating. For \textit{(i)} and \textit{(ii)}, we use the analytical fit of \cite{Rollig06} to the exact multilevel treatment by \cite{Sternberg95}. For \textit{(iii)}, we follow \cite{Kamp01} and assume that one third of the binding energy of $\sim 4.5$ eV is released to the gas for heating. Photodissociation \textit{(iv)} carries away about 0.4 eV in the form of heat to the gas (e.g. \citealt{Jonkheid04}). We note that the H$_2$ dissociation rate is taken to be consistent with the chemistry.
 
\subsubsection*{4.) Cosmic-ray heating}

Following \cite{Cravens78} and \cite{Glassgold73}, a cosmic-ray ionization of H$_2$ (H) releases 8 eV (3.5 eV) to the gas.

\subsubsection*{5.) X-ray heating}

We implement the X-ray heating rates following \cite{Dalgarno99} for a H$_2$, H, and He mixture. The energy deposition rate per particle H$_{\rm x}$ is calculated using the cross-sections provided in \cite{Bruderer09a}.

\subsubsection*{6.) Line cooling (molecular and atomic fine-structure lines)}

Cooling rates of molecular lines and atomic fine-structure lines are derived from the level population calculation (Sect. \ref{sec:modline}). We account for O, C, C$^+$, CO, ${}^{13}$CO, H$_2$O, and OH. The abundance of ${}^{13}$CO is obtained from scaling the CO abundance by ${}^{12}{\rm C} /{}^{13}{\rm C} = 70$ (\citealt{Wilson94}). Molecular data are taken from the LAMDA database (\citealt{Schoier05}) and include the collisional excitation with H, H$_2$, and electrons for O, C, and C$^+$, respectively. For the molecular species, only excitation with H$_2$ is available. 

\subsubsection*{7.) Hydrogen Ly$\alpha$ and [\ion{O}{I}] 6300 \AA~ cooling}

Hydrogen Ly$\alpha$ cooling and cooling by neutral oxygen [\ion{O}{I}] by means of the metastable ${}^1$D-${}^3$P line at 6300 \AA~ is included following \cite{Sternberg89}.

\subsubsection*{8.) Carbon ionization}

Carbon ionization delivers about 1 eV to the gas (\citealt{Jonkheid04}). The ionization rate is assumed to be consistent with the chemical network including the effects of self-shielding.

\subsubsection{Benchmark test}

\begin{figure*}[tbh]
\centering
\includegraphics[width=1.0\hsize]{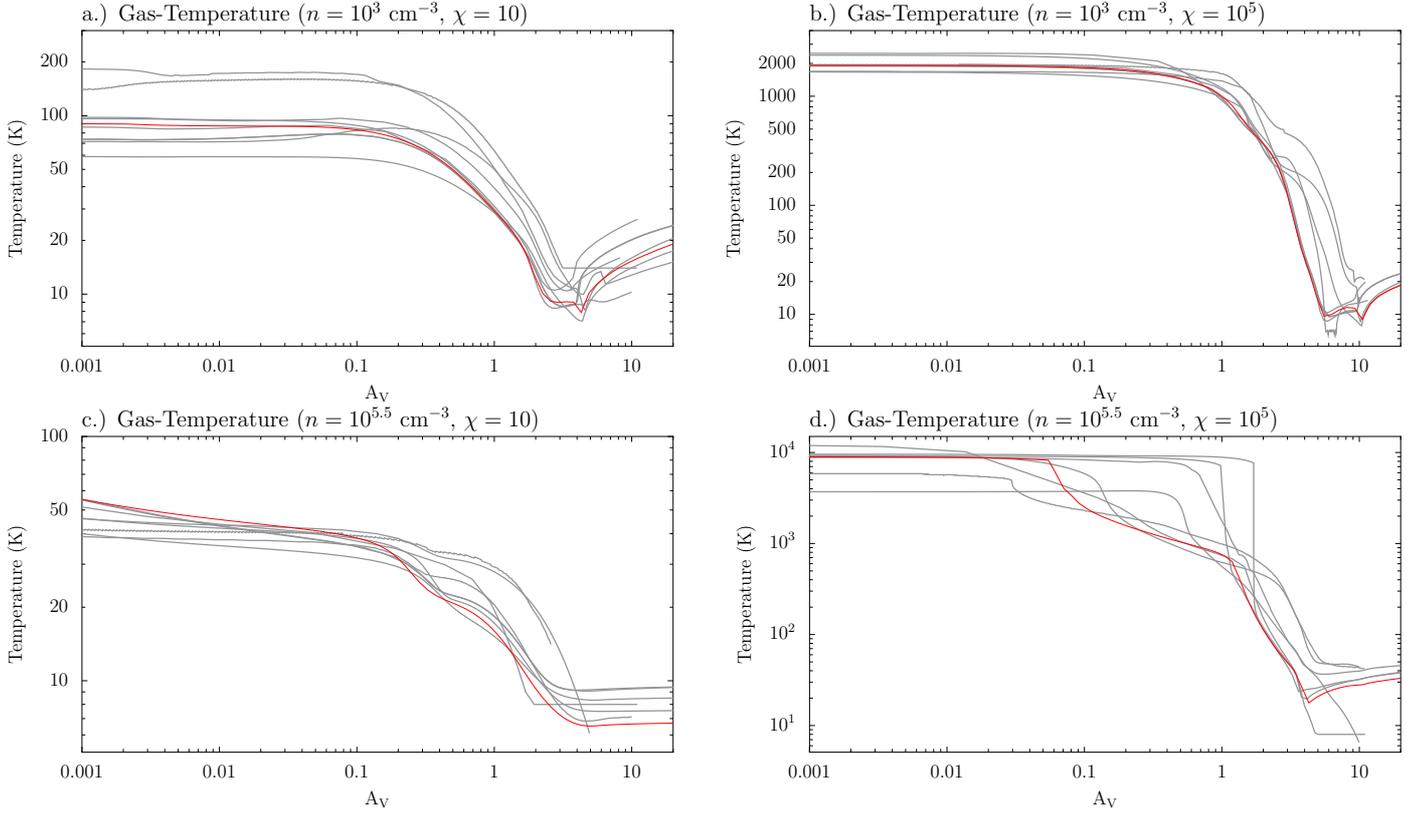}
\caption{Result of benchmark test for the calculation of the gas temperature. The results of the codes participating in the study by \cite{Roellig07} are shown in gray, and the gas temperature obtained with our code is given in red. \label{fig:benchmark_temp}}
\end{figure*}

To test the implementation of the thermal balance calculation, we run the benchmark problem of the PDR comparison study of \cite{Roellig07}, which consists of four slab models with a density of $10^{3}$ and $10^{5.5}$ cm$^{-3}$ and an incident UV radiation field of $\chi=10$ or $\chi=10^5$ (in units of Draine ISRF, \citealt{Draine78}). To eliminate the  differences caused by the adopted chemical network, the benchmark study is run with the same simple chemical network consisting of only the 31 species used in the PDR comparison study. The parameters given in Table 5 of \cite{Roellig07} are implemented.

The gas temperature calculated with our code is shown in Fig. \ref{fig:benchmark_temp}, together with the results from other PDR models. The gas temperature derived with our code is in reasonable agreement with that found by other PDR codes. Differences between the different codes are however considerable, particularly for the high-density ($10^{5.5}$ cm$^{-3}$) model with high UV intensity ($\chi=10^5$). This model, is however closest to the conditions in disk atmospheres and outflow walls.

\section{Analytical estimates of fluxes}

\subsection{Analytical estimate of the [\ion{C}{II}] flux} \label{sec:app_cplus}

To analytically estimate the [\ion{C}{II}] flux that might originate from a remnant envelope or the foreground, we use 
\begin{equation}
F=d\Omega \cdot I = \frac{A}{d^2} \frac{h\nu}{4\pi} A_{ul} N_{{\rm C}^+} x_u \ ,
\end{equation}
for the flux $F$, the integrated intensity $I$, and the solid angle $d\Omega$. The solid angle of the emitting area is calculated from the projected area $A$ and the distance $d$ to the source. The integrated intensity $I$ is obtained from the line frequency $\nu$, the Einstein-A coefficient $A_{ul}$, the column density of C$^+$ $N_{{\rm C}^+}$, and the normalized level in the upper level $x_u$. Assuming the level population in the LTE, which is the case above the critical density of a few times $10^3$ cm$^{-3}$, we reach a maximum $x_u=2/3$, given by the statistical weights of the C$^+$ levels. These equations assume optically thin emission. For a line width of 1 km s$^{-1}$, $\tau=1$ at the line center is reached for a column density of a few times $10^{18}$ cm$^{-2}$.

\subsection{Rotation diagram of CO} \label{sec:app_rotco}

For the rotation diagram of CO, we consider the optically thin flux of CO, which is assumed to be thermalized,
\begin{equation} \label{eq:rotbas}
F_{ul} = d\Omega \cdot I_{ul} = d\Omega \frac{h\nu_{ul}}{4\pi} A_{ul} \, N({\rm CO}) \frac{g_u e^{-E_u/k T}}{Q(T)} \,
\end{equation}
using the same notation as in Appendix \ref{sec:app_cplus} and the CO column density $N({\rm CO})$, the upper level energy $E_u$, and the partition function $Q(T)$. Rearranging yields
\begin{equation}
e^Y \equiv \frac{4 \pi F_{ul}}{A_{ul} h\nu_{ul} g_u} = d\Omega N({\rm CO}) \frac{e^{-E_u/k T}}{Q(T)} \equiv d\Omega \frac{N_u}{g_u} \ ,
\end{equation}
with the column density per level $N_u = N({\rm CO}) \frac{g_u e^{-E_u/k T}}{Q(T)}$. Thus,
\begin{equation}
Y= \ln\left(\frac{4 \pi F_{ul}}{A_{ul} h\nu_{ul} g_u}\right) = \ln\left( d\Omega \frac{N({\rm CO})}{Q(T)}\right) - \frac{E_u}{k T} \ .
\end{equation}
We note that the rotation diagram is often given by assuming that $F_{ul} = d\Omega_{\rm beam} I_{ul}$, for the solid angle of the telescope beam $d\Omega_{\rm }$, thus the molecules are equally distributed over the beam. This is however inappropriate for the CO ladder probing a wide range of temperatures and subsequently different emitting regions (Section \ref{sec:lineorigin}).

How do opacity effects alter the rotation diagram? Assuming a square line profile of (velocity) width $\Delta v$ for simplicity and neglecting the background intensity, we obtain for the integrated intensity using Eq. \ref{eq:linert_solveradi}),
\begin{equation} \label{eq:iint_full}
I_{ul} = \Delta v \frac{\nu_{ul}}{c} B(T_{{\rm ex},ul}) ( 1 - e^{-\tau_{ul}}) \ .
\end{equation}
For $\tau_{ul} \ll 1$, $I_{ul} \sim \Delta v \frac{\nu_{ul}}{c} B(T_{{\rm ex},ul}) \tau_{ul}= j_\nu dl$, recovering the expression used in Eq. \ref{eq:rotbas}. For $\tau_{ul} > 1$, $I_{ul} \sim \Delta v \frac{\nu_{ul}}{c} B(T_{{\rm ex},ul})$, which is always larger than the optically thin expression. Thus, optical depth effects shift points above the optically thin lines in the rotation diagram.

\end{appendix}

\end{document}